
\newif\iffigs\figstrue

\input harvmac
\iffigs
  \input epsf
\else
  \message{No figures will be included.  See TeX file for more
information.}
\fi

%
%
\message{S-Tables Macro v1.0, ACS, TAMU (RANHELP@VENUS.TAMU.EDU)}
%
%
\newhelp\stablestylehelp{You must choose a style between 0 and 3.}%
\newhelp\stablelinehelp{You
should not use special hrules when stretching
a table.}%
\newhelp\stablesmultiplehelp{You have tried to place an S-Table
inside another
S-Table.  I would recommend not going on.}%
%
%
\newdimen\stablesthinline
\stablesthinline=0.4pt
\newdimen\stablesthickline
\stablesthickline=1pt
%
%
\newif\ifstablesborderthin
\stablesborderthinfalse
\newif\ifstablesinternalthin
\stablesinternalthintrue
\newif\ifstablesomit
\newif\ifstablemode
\newif\ifstablesright
\stablesrightfalse
%
%
\newdimen\stablesbaselineskip
\newdimen\stableslineskip
\newdimen\stableslineskiplimit
%
%
\newcount\stablesmode
\newcount\stableslines
\newcount\stablestemp
\stablestemp=3
\newcount\stablescount
\stablescount=0
\newcount\stableslinet
\stableslinet=0
%
%
%
\newcount\stablestyle
\stablestyle=0
%
%
\def\stablesleft{\quad\hfil}%
\def\stablesright{\hfil\quad}%
%
%
\catcode`\|=\active%
%
%
\newcount\stablestrutsize
\newbox\stablestrutbox
\setbox\stablestrutbox=\hbox{\vrule height10pt depth5pt width0pt}
\def\stablestrut{\relax\ifmmode%
                         \copy\stablestrutbox%
                       \else%
                         \unhcopy\stablestrutbox%
                       \fi}%
%
%
\newdimen\stablesborderwidth
\newdimen\stablesinternalwidth
\newdimen\stablesdummy
\newcount\stablesdummyc
\newif\ifstablesin
\stablesinfalse
%
%
\def\begintable{\stablestart%
  \stablemodetrue%
  \stablesadj%
  \halign%
  \stablesdef}%
\def\stablesadj{%
  \ifcase\stablestyle%
    \hbox to \hsize\bgroup\hss\vbox\bgroup%
  \or%
    \hbox to \hsize\bgroup\vbox\bgroup%
  \or%
    \hbox to \hsize\bgroup\hss\vbox\bgroup%
  \or%
    \hbox\bgroup\vbox\bgroup%
  \else%
    \errhelp=\stablestylehelp%
    \errmessage{Invalid style selected, using default}%
    \hbox to \hsize\bgroup\hss\vbox\bgroup%
  \fi}%
\def\stablesend{\egroup%
  \ifcase\stablestyle%
    \hss\egroup%
  \or%
    \hss\egroup%
  \or%
    \egroup%
  \or%
    \egroup%
  \else%
    \hss\egroup%
  \fi}%
\def\stablestart{%
  \ifstablesin%
    \errhelp=\stablesmultiplehelp%
    \errmessage{An S-Table cannot be placed within an S-Table!}%
  \fi
  \global\stablesintrue%
  \global\advance\stablescount by 1%
  \message{<S-Tables Generating Table \number\stablescount}%
  \begingroup%
  \stablestrutsize=\ht\stablestrutbox%
  \advance\stablestrutsize by \dp\stablestrutbox%
  \ifstablesborderthin%
    \stablesborderwidth=\stablesthinline%
  \else%
    \stablesborderwidth=\stablesthickline%
  \fi%
  \ifstablesinternalthin%
    \stablesinternalwidth=\stablesthinline%
  \else%
    \stablesinternalwidth=\stablesthickline%
  \fi%
  \tabskip=0pt%
  \stablesbaselineskip=\baselineskip%
  \stableslineskip=\lineskip%
  \stableslineskiplimit=\lineskiplimit%
  \offinterlineskip%
  \def\borderrule{\vrule width \stablesborderwidth}%
  \def\internalrule{\vrule width \stablesinternalwidth}%
  \def\thinline{\noalign{\hrule height \stablesthinline}}%
  \def\thickline{\noalign{\hrule height \stablesthickline}}%
  \def\trule{\omit\leaders\hrule height \stablesthinline\hfill}%
  \def\ttrule{\omit\leaders\hrule height \stablesthickline\hfill}%
  \def\tttrule##1{\omit\leaders\hrule height ##1\hfill}%
  \def\stablesel{&\omit\global\stablesmode=0%
    \global\advance\stableslines by 1\borderrule\hfil\cr}%
  \def\el{\stablesel&}%
  \def\elt{\stablesel\thinline&}%
  \def\eltt{\stablesel\thickline&}%
  \def\elttt##1{\stablesel\noalign{\hrule height ##1}&}%
  \def\elspec{&\omit\hfil\borderrule\cr\omit\borderrule&%
              \ifstablemode%
              \else%
                \errhelp=\stablelinehelp%
                \errmessage{Special ruling will not display properly}%
              \fi}%
  \def\stmultispan##1{\mscount=##1 \loop\ifnum\mscount>3
\stspan\repeat}%
  \def\stspan{\span\omit \advance\mscount by -1}%
  \def\multicolumn##1{\omit\multiply\stablestemp by ##1%
     \stmultispan{\stablestemp}%
     \advance\stablesmode by ##1%
     \advance\stablesmode by -1%
     \stablestemp=3}%
  \def\multirow##1{\stablesdummyc=##1\parindent=0pt\setbox0\hbox\bgroup%
    \aftergroup\emultirow\let\temp=}
  \def\emultirow{\setbox1\vbox to\stablesdummyc\stablestrutsize%
    {\hsize\wd0\vfil\box0\vfil}%
    \ht1=\ht\stablestrutbox%
    \dp1=\dp\stablestrutbox%
    \box1}%

\def\stpar##1{\vtop\bgroup\hsize ##1%
     \baselineskip=\stablesbaselineskip%
     \lineskip=\stableslineskip%

\lineskiplimit=\stableslineskiplimit\bgroup\aftergroup\estpar\let\temp=}%
  \def\estpar{\vskip 6pt\egroup}%
  \def\stparrow##1##2{\stablesdummy=##2%
     \setbox0=\vtop to ##1\stablestrutsize\bgroup%
     \hsize\stablesdummy%
     \baselineskip=\stablesbaselineskip%
     \lineskip=\stableslineskip%
     \lineskiplimit=\stableslineskiplimit%
     \bgroup\vfil\aftergroup\estparrow%
     \let\temp=}%
  \def\estparrow{\vfil\egroup%
     \ht0=\ht\stablestrutbox%
     \dp0=\dp\stablestrutbox%
     \wd0=\stablesdummy%
     \box0}%
  \def|{\global\advance\stablesmode by 1&&&}%
  \def\|{\global\advance\stablesmode by 1&\omit\vrule width 0pt%
         \hfil&&}%
  \def\vt{\global\advance\stablesmode by 1&\omit\vrule width
\stablesthinline%
          \hfil&&}%
  \def\vtt{\global\advance\stablesmode by 1&\omit\vrule width
\stablesthickline%
          \hfil&&}%
  \def\vttt##1{\global\advance\stablesmode by 1&\omit\vrule width ##1%
          \hfil&&}%
  \def\vtr{\global\advance\stablesmode by 1&\omit\hfil\vrule width%
           \stablesthinline&&}%
  \def\vttr{\global\advance\stablesmode by 1&\omit\hfil\vrule width%
            \stablesthickline&&}%
  \def\vtttr##1{\global\advance\stablesmode by 1&\omit\hfil\vrule
width ##1&&}%
  \stableslines=0%
  \stablesomitfalse}
\def\stablesdef{\bgroup\stablestrut\borderrule##\tabskip=0pt plus 1fil%
  &\stablesleft##\stablesright%
  &##\ifstablesright\hfill\fi\internalrule\ifstablesright\else\hfill\fi%
  \tabskip 0pt&&##\hfil\tabskip=0pt plus 1fil%
  &\stablesleft##\stablesright%
  &##\ifstablesright\hfill\fi\internalrule\ifstablesright\else\hfill\fi%
  \tabskip=0pt\cr%
  \ifstablesborderthin%
    \thinline%
  \else%
    \thickline%
  \fi&%
}%
\def\endtable{\advance\stableslines by 1\advance\stablesmode by 1%
   \message{- Rows: \number\stableslines, Columns:
\number\stablesmode>}%
   \stablesel%
   \ifstablesborderthin%
     \thinline%
   \else%
     \thickline%
   \fi%
   \egroup\stablesend%
\endgroup%
\global\stablesinfalse}
%

\catcode`\|=12

\noblackbox
\input epsf
\def\inbar{\vrule height1.5ex width.4pt depth0pt}
\def\IC{\relax\hbox{\kern.25em$\inbar\kern-.3em{\rm C}$}}
\def\IP{\relax{\rm I\kern-.18em P}}
\def\IF{\relax{\rm I\kern-.18em F}}
\def\IZ{\relax\ifmmode\hbox{Z\kern-.4em Z}\else{Z\kern-.4em Z}\fi}
\def\IR{\relax{\rm I\kern-.18em R}}
\def\ICP{{\IC\IP}}

\Title{\vbox{\baselineskip14pt
\hbox{CU-TP-769}\hbox{DUKE-TH-96-125}\hbox{HUTP-96/A033}\hbox{hep-th/9608039}}}
{A Geometric Realization of Confinement}
\centerline{Brian R. Greene\footnote{$^*$}{On leave from: F.R. Newman
Laboratory,
Cornell University, Ithaca, NY 14853}}
\smallskip
\centerline{\it  Departments of Mathematics and Physics, Columbia University}
\centerline{\it New York, NY 10027, USA}
\bigskip
\centerline{David R. Morrison}
\smallskip
\centerline{\it Department of Mathematics, Duke University}
\centerline{\it Durham, NC 27708, USA}
\bigskip\centerline{and}\smallskip
\centerline{Cumrun Vafa}
\smallskip
\centerline{\it  Lyman Laboratory of Physics, Harvard
University}
\centerline{\it Cambridge, MA 02138, USA}
\vskip .3in
We study the geometric realization of the Higgs phenomenon
in type II string compactifications on Calabi--Yau manifolds.
The string description is most directly phrased in terms of confinement
of magnetic flux, with magnetic charged states arising
from D-branes wrapped around chains
as opposed to cycles.  The rest of the closed cycle of the
D-brane worldvolume is manifested as a confining flux tube emanating
from the magnetic charges, in the uncompactified space.
We also study corrections to hypermultiplet
moduli for type II compactifications, in particular for
type IIA near the conifold point.
\Date{\it {August, 1996}}

\lref\lefschetz{S. Lefschetz, L'Analysis Situs et la G\'eom\'etrie
Alg\'ebrique, Gauthier-Villars, Paris, 1924; reprinted in
Selected Papers, Chelsea, New York, 1971, pp.~283--439.}
\lref\rQT{F. Quevedo and C. A. Trugenberger, {\it Phases of Antisymmetric
Tensor Field Theories}, {\tt hep-th/9604196}.}
\lref\rBerkSie{N. Berkovits and W. Siegel, {\it Superspace Effective
Actions for 4D Compactifications of Heterotic and Type II Superstrings},
Nucl. Phys. {\bf B462} (1996) 213--248,
{\tt hep-th/9510106}.}
\lref\rKleb{A. Hanany and I. R. Klebanov, {\it On Tensionless Strings
in $3+1$ Dimensions}, {\tt hep-th/9606136}.}
\lref\rNielsen{Nielsen--Olesen, Cosmic Strings}
\lref\rKron{P. B. Kronheimer, {\it The Construction of ALE Spaces as
Hyper-K\"ahler Quotients}, J. Differential Geom. {\bf 29} (1989) 665--683.}
\lref\vf{C. Vafa, {\it Evidence for F-Theory}, Nucl. Phys.
{\bf B469} (1996) 403--418, {\tt hep-th/9602022}.}
\lref\scs{B. R. Greene, A. Shapere, C. Vafa, and S.-T. Yau,
     {\it Stringy Cosmic Strings}, Nucl. Phys. {\bf B337} (1990) 1--36.}
\lref\ss{N. Seiberg and A. Strominger, unpublished.}
\lref\dlp{
J.~Dai, R.~Leigh, and J.~Polchinski, {\it New Connections Between String
  Theories}, Mod. Phys. Lett. A {\bf 4} (1989) 2073--2083.}
\lref\witb{E. Witten, {\it Bound States of Strings and $p$-Branes},
     {\tt hep-th/9510135}.}
\lref\fer{
S.~Ferrara and S.~Sabharwal, {\it Quaternionic Manifolds for Type {II}
  Superstring Vacua of {C}alabi--{Y}au Spaces}, Nucl. Phys. B {\bf 332} (1990)
  317--332.}
\lref\CFG{
S.~Cecotti, S.~Ferrara, and L.~Girardello, {\it Geometry of Type {II}
  Superstrings and the Moduli of Superconformal Field Theories}, Int. J. Mod.
  Phys. A {\bf 4} (1989) 2475--2529.}

\lref\sew{N. Seiberg and E. Witten,
{\it Electric-Magnetic Duality,
     Monopole Condensation and Confinement in $N{=}2$ Supersymmetric
     Yang--Mills Theory},
     Nucl. Phys. {\bf B426} (1994) 19--52, {\tt hep-th/9407087}.}

\lref\kv{S. Kachru and C. Vafa, {\it Exact Results for $N{=}2$
     Compactifications of Heterotic Strings}, Nucl. Phys. {\bf B450} (1995)
     69--89, {\tt hep-th/9505105}.}
\lref\kklmv{S. Kachru,  A. Klemm, W. Lerche, P. Mayr and C. Vafa,
     {\it Non-Perturbative Results on the Point Particle Limit
     of $N{=}2$ Heterotic String Compactifications}, Nucl. Phys. {\bf B459}
     (1996) 537--588, {\tt hep-th/9508155}.}
\lref\swt{N. Seiberg and E. Witten, {\it Gauge Dynamics and Compactification
to Three Dimensions}, {\tt hep-th/9607163}.}
\lref\ov{H. Ooguri and C. Vafa, {\it Two-Dimensional Black Hole and
Singularities of CY Manifolds}, Nucl. Phys. {\bf B463} (1996) 55--72,
{\tt hep-th/9511164}.}
\lref\bsvii{M. Bershadsky, V. Sadov, and C. Vafa, {\it D-Strings on
D-Manifolds},
Nucl. Phys. {\bf B463} (1996) 398--414, {\tt hep-th/9510225}}
\lref\bsv{M. Bershadsky, V. Sadov, and C. Vafa, {\it
     $D$-Branes and Topological Field Theories}, Nucl. Phys. {\bf B463} (1996)
420--434,
     {\tt hep-th/9511222}.}
\lref\klmvw{A. Klemm, W. Lerche, P. Mayr, C. Vafa and N. Warner,
{\it Self-Dual Strings and $N{=}2$ Supersymmetric Field Theory},
{\tt hep-th/9604034}.}
\lref\rBBS{K. Becker, M. Becker, and A. Strominger, {\it
     Fivebranes, Membranes and Non-Per\-tur\-ba\-tive String Theory},
     Nucl. Phys. {\bf B456} (1995) 130--152,
     {\tt hep-th/9507158}.}
\lref\rHL{
R.~Harvey and H.~B. Lawson, Jr., {\it Calibrated Geometries}, Acta Math. {\bf
  148} (1982) 47--157.}
\lref\rMcLean{
R.~C. McLean, {\it Deformations of Calibrated Submanifolds},
 J. Differential Geom., to appear, preprint
   at {\tt http://www.math.duke.edu/preprints/1996.html}.}
\lref\rAGMsmall{
     P. S. Aspinwall, B. R. Greene and D. R. Morrison, {\it Measuring Small
     Distances in
     $N{=}2$ Sigma Models}, Nucl. Phys.  {\bf B420} (1994) 184--242,
     {\tt hep-th/9311042}.}
\lref\rBAT{V. V. Batyrev, {\it Dual Polyhedra and Mirror Symmetry for
     Calabi--Yau Hypersurfaces in Toric Varieties}, J. Alg. Geom. {\bf 3}
     (1994) 493--535, {\tt alg-geom/9310003}.}
\lref\rAGMmultiple{P. S. Aspinwall, B. R. Greene and D. R. Morrison,
{\it Multiple Mirror Manifolds and Topology Change in String Theory},
Phys. Lett. {\bf B303} (1993) 249--259, {\tt hep-th/9301043}.}
\lref\rAGMmath{P. S. Aspinwall, B. R. Greene, and D. R. Morrison, {\it
     The Monomial-Divisor Mirror Map}, Internat. Math. Res. Notices
     (1993) 319--337, {\tt alg-geom/9309007}.}
\lref\rWittenphases{E. Witten, {\it Phases of $N{=}2$ Theories In Two
Dimensions},
     Nucl. Phys. {\bf B403} (1993) 159--222, {\tt hep-th/9301042}.}
\lref\rAGM{P. S. Aspinwall, B. R. Greene, and D. R. Morrison, {\it
     Calabi--Yau Moduli Space, Mirror Manifolds and Spacetime Topology
     Change in String Theory}, Nucl. Phys. {\bf B416} (1994) 414--480,
     {\tt hep-th/9309097}.}
\lref\rPolch{J. Polchinksi, {\it Dirichlet Branes and Ramond-Ramond Charges},
Phys. Rev. Lett. {\bf 75} (1995) 4724--4727, {\tt hep-th/9510017}.}
\lref\rStrominger{A. Strominger, {\it Massless Black Holes and Conifolds in
String
     Theory}, Nucl. Phys. {\bf B451} (1995) 97--109, {\tt hep-th/9504090}.}
\lref\rGMS{B. R. Greene, D. R. Morrison, and A. Strominger,
     {\it Black Hole Condensation and the Unification of String Vacua},
     Nucl. Phys. {\bf B451} (1995) 109--120, {\tt hep-th/9504145}.}
\lref\rWittenGauge{E. Witten, {\it String Theory Dynamics in Various
Dimensions},
Nucl. Phys. {\bf B443} (1995) 85--126, {\tt hep-th/9503124}.}

\lref\clemens{C. H. Clemens,
{\it Double Solids},  Adv. Math. {\bf 47} (1983) 107--230.}
\lref\friedman{R. Friedman,
{\it Simultaneous Resolution of Threefold Double Points},
Math. Ann. {\bf 274} (1986) 671--689.}
\lref\hirzebruch{F. Hirzebruch (notes by J. Werner),
{\it Some Examples of Threefolds
with Trivial Canonical Bundle}, in Gesammelte Abhandlungen, Band II,
Springer-Verlag, Berlin-New York, 1987, pp.~757--770.}
\lref\tianyau{G. Tian and S.-T. Yau,
{\it Three Dimensional Algebraic Manifolds with $c_1=0$ and $\chi=-6$},
in Mathematical Aspects of String Theory (S.-T. Yau, ed.),
World Scientific, Singapore, 1987, pp.~629--646.}
\lref\texasi{ P. Candelas, A. M. Dale, C. A. L\"utken, and R. Schimmrigk,
{\it Complete Intersection Calabi--Yau Manifolds}, Nucl. Phys.
{\bf B298} (1988) 493--525.}
\lref\texasii{P. S. Green and T.  H\"ubsch,
{\it Possible Phase Transitions Among Calabi--Yau Compactifications},
Phys. Rev. Lett.  {\bf 61} (1988) 1163--1166;
{\it Connecting Moduli Spaces of Calabi--Yau Threefolds}, Comm. Math.
Phys.  {\bf 119} (1988) 431--441.}
\lref\texasiii{P. Candelas, P. S. Green, and T. H\"ubsch,
{\it Finite Distance Between Distinct Calabi--Yau Manifolds},
Phys. Rev. Lett.  {\bf 62} (1989) 1956--1959;
{\it Rolling Among Calabi--Yau Vacua}, Nucl. Phys.
{\bf B330} (1990) 49--102.}
\lref\cdcon{ P. Candelas and X. C. de la Ossa,
{\it Comments on Conifolds}, Nucl. Phys. {\bf B342} (1990) 246--268.}
\lref\rTrento{D. R. Morrison, ``Through the Looking Glass'',
Lecture at CIRM conference, Trento (June, 1994), to appear.}

\newsec{Introduction}

A tremendous amount of progress has been made
recently in understanding nonperturbative
aspects of string theory. One central tool
in these developments has been the window
on nonperturbative physics opened by the study
of solitonic states and their subsequent microscopic
description in terms of D-branes
\rPolch. In compactified string theory,
a by now familiar application of these ideas is to consider
the physics associated with wrapping various
$p$-branes around nontrivial homology cycles in
spacetime. For instance, such considerations have
led to an understanding of enhanced gauge symmetry \rWittenGauge,
conifold singularities
\rStrominger, topology-changing conifold transitions
\rGMS,
and many other striking developments.

In this paper we initiate a study which is a natural
outgrowth of the above considerations.  In particular,
attempts to describe the inverse of the transition
studied in \rGMS, i.e. a geometric
description of un-Higgsing, naturally leads us to
consider  D-branes which wrap around
 {\it chains}\/
as opposed to cycles. That is, we wrap D-branes
around submanifolds with nontrivial boundaries.
We find that they correspond to magnetically charged particles
in the confining phase.  Just as a free quark in the confining
phase does not make sense in isolation, D-branes wrapped
around chains do not make sense in isolation either. However, we can
consider pairs of such magnetically charged objects, separated
in the uncompactified spacetime, and we find that  pairs
 of such chains can be assembled into
cycles whose image in spacetime  includes magnetic flux
tubes.
In this way, although we mainly
consider the abelian case here, we find
a simple geometric interpretation of gauge
confinement which we believe has bearing on a geometric
interpretation of confinement even in the non-abelian case.

 In \rGMS\ conifold transitions
in type II string theory were found
by moving to a singular
point in the vector multiplet moduli space and
passing on to a new Higgs branch by giving appropriate
vacuum expectation values to new massless hypermultiplets
which arise. This
corresponds geometrically to degenerating the
complex structure of a Calabi--Yau in the type IIB theory
 and then performing a small resolution, or,
by using mirror symmetry, to
degenerating the K\"ahler structure of a Calabi--Yau
in type IIA and then performing a desingularizing
deformation. These descriptions, though, only
cover half of the story. Abstractly, one can wonder
about the ``reverse'' processes in which one first
moves in the hypermultiplet moduli space and then
finds a new Coulomb branch. For ease of
reference, we will call the transitions
studied in \rGMS\ {\it vector}-conifold
transitions and the reverse process
 {\it hyper}-conifold transitions, where the prefix indicates
the type of parameters initially being varied.
 Concretely, hyper-conifold transitions arise by
initially degenerating the K\"ahler
structure of a type IIB string or, by mirror
symmetry, the complex structure of a type IIA string.
It is not a trivial change of perspective to consider
these reverse processes from \rGMS\ for two reasons.
First, whereas the vector multiplet moduli space
is governed by special geometry and receives no quantum corrections,
we know far less about
the quantum-corrected quaternionic geometry of the hypermultiplet
moduli space. Second, type IIA theory has
even-dimensional branes while type IIB has odd-dimensional
branes.
In \rGMS, odd-cycles were degenerated in the type IIB
context (equivalently, even-cycles were degenerated
in the type IIA context) and hence a local
description in terms of {\it particles}\/ was
obtained by wrapping the appropriate odd (even)
branes around the vanishing cycles. In hyper-conifold
processes, even (odd) cycles are degenerating in
the type IIB (IIA) context and hence we, at first
sight, lose a particle description and are forced
to understand some of the physics of tensionless
strings. In the present context, we unravel the physics
of these tensionless strings and reinterpret {\it pairs}\/ of them in terms of
magnetically confined flux tubes.
 We will see that a key part is played
by wrapping branes on chains, giving rise to magnetically charged
states that are linked together via such nearly-tensionless strings.
The fact that magnetically charged objects are linked by confining
flux tubes is reflected in the geometric fact that chains must be
attached to other chains in order to make a closed three-cycles on which branes
can be wrapped.

In section 2 we explain the geometric description
of conifold transitions in somewhat greater
detail then was done in \rGMS\ as this is
required for our present study. In section 3
we discuss some simple field theory phenomena
whose string theory counterparts comprise
section 4. In the latter section we begin
by following the fate of massive {\it magnetic}\/
solitons through the vector-conifold transitions
of \rGMS. This naturally yields magnetic
confinement as the physical interpretation of
our mathematical description in section 2.
We develop this picture in some detail which leads us
to a deeper
understanding of the reverse process of
hyper-conifold transitions.  Moreover in this
geometric approach to confinement we find that the confined
charge and the confined flux tube are unified
into a single geometric object in higher dimension whose projection
in spacetime leads to an apparent asymmetry between the
charge and the flux tube.
 In section 5 we
elaborate on \rGMS\ and on section 2 by
delineating, at the level of conformal
field theory, the location in hypermultiplet
moduli space where these hyper-conifold transitions
occur and therefore clarify the attachment of
the associated Coulomb branch. In section 6 we
describe some aspects of quantum string corrections
for hypermultiplet moduli motivated from the description
of the above transition.  Finally in section 7 we
end with some conclusions and speculations.

\newsec{The Geometry of Conifold Transitions}

We begin by reviewing the mathematics of conifold transitions,
following both the original mathematics literature
\refs{\clemens\friedman\hirzebruch{--}\tianyau},
and subsequent discussions of these transitions in the physics literature
\refs{\texasi\texasii\texasiii{--}\cdcon}.  The transition can be
approached by varying either the complex structure or the K\"ahler structure,
and we shall discuss both perspectives.

The set of possible complex structures on a given Calabi--Yau
threefold can be parameterized, in
favorable circumstances, by the coefficients in the defining
equation or equations of that threefold.  A familiar example of this is the
case of the quintic hypersurface in $\ICP^4$, in which 101 of the coefficients
of the quintic polynomial can be used as parameters.  At general values
of these parameters, the solution set describes a smooth manifold, but
at special values, the solution set will have singularities.

The simplest singularities which one encounters are {\it nodes}, that is,
singularities which can be locally described at the singular point
by an equation whose constant and linear terms at that point
 vanish, and whose quadratic
terms define a quadratic of maximal rank.  This phenomenon---of acquiring a
single node---happens at
(complex) codimension one in the complex structure moduli space, since there is
one algebraic condition on the parameters which characterizes when the
solution set fails to be nonsingular.  The set of points where a node
has been acquired is called the {\it conifold locus}\/ in the moduli space.

Associated to any node which is acquired in this fashion
is a so-called {\it vanishing cycle}.  This is a three-cycle which
exists on the nearby nonsingular Calabi--Yau manifolds, and which can
be seen topologically by intersecting the Calabi--Yau manifolds with
a small ball in the ambient space which surrounds the singular point.
The boundary of that intersection (i.e. the intersection with a small sphere)
will have topology $S^2\times S^3$, and in fact the singular Calabi--Yau
space looks topologically like a real cone over $S^2\times S^3$ near the
singular point.  However, in the nearby nonsingular Calabi--Yau manifolds,
the $S^2\times S^3$ will be the boundary of some $B^3\times S^3$, which is
a neighborhood of the ``vanishing cycle'' $0\times S^3$.

If we call the homology class of the
vanishing cycle $\gamma_1$, then it is possible to find a basis
$\gamma_0, \gamma_1, \dots, \gamma_k$ for a Lagrangian subspace of
$H_3(X,\IZ)$ and a holomorphic
three-form $\Omega$ so that
$\int_{\gamma_0}\Omega$ is single-valued and non-zero near the conifold locus,
and the corresponding periods $Z^1,\dots,Z^k$ defined by
$Z^j=\int_{\gamma_j}\Omega$
give multi-valued
local coordinates on the complex structure moduli space near the conifold
locus.  The period $Z^1=\int_{\gamma_1}\Omega$ is zero along the entire
conifold locus,
which is why $\gamma_1$ is called a vanishing cycle.

The vanishing cycle is expected to have a unique representation as a
supersymmetric three-cycle \rBBS\ (i.e., there should be a unique ``special
Lagrangian submanifold'' \refs{\rHL,\rMcLean} in this homology
class near the node).
Since for such a representative, the volume is proportional to the absolute
value of the period,
the supersymmetric vanishing cycle  gets smaller and smaller as the conifold
locus is approached, and the cycle literally vanishes in the limit.

In the conifold transition, the singularity of the limiting Calabi--Yau
space is resolved by a type of blowing up, in which the node is replaced
 by a holomorphically
embedded $S^2$. (This is known as a {\it small resolution}\/ of the
node.) Topologically, the new $S^2$ has a neighborhood of the form
$S^2\times B^4$ whose boundary is $S^2\times S^3$:
this is the same $S^2\times S^3$
(up to homotopy) as was obtained by intersection with a ball in the ambient
space.

The space which one obtains by blowing up the singularity
in this way is always a complex manifold, but
it will fail to be a K\"ahler manifold if only a single node is involved,
as we will see later.
In fact, even if we have a Calabi--Yau space with several nodes,
the condition for the existence of a K\"ahler metric after blowing
up these nodes is a somewhat subtle one:  the vanishing cycle of each
node must be homologous to a nonzero linear combination of vanishing cycles
of the other nodes.\foot{These homology relations can be represented by
four-chains whose boundaries are linear combinations of vanishing cycles; in
the limit, these four-chains become four-cycles which can actually be
represented by complex submanifolds (of complex codimension one).
Blowing up those  submanifolds
resolves the singularities, with the blown up space naturally embedded in
$X\times \ICP^1\times \dots \times \ICP^1$; it inherits a K\"ahler metric
from that of the ambient space.}

The areas of the holomorphic two-spheres are new classes in cohomology
after the blowup (although not all of them will be independent).  In fact, the
reverse of the blowing up process can be regarded as a variation of the
K\"ahler parameters in which all of the areas of the holomorphic two-spheres
are sent to zero, producing again the singular Calabi--Yau space.
Thus, when we approach the transition from one side we see vanishing
supersymmetric three-spheres, and when we approach it from the other side we
see shrinking holomorphic two-spheres.

For example, if we vary the defining equation of a quintic hypersurface in
$\ICP^4$ until it contains some fixed $\ICP^2$, the limiting quintic has
sixteen nodes along the $\ICP^2$.  If we remove from $\ICP^2$ small balls
around
each of the nodes, we find a four-chain whose boundary is the sum of the
sixteen vanishing three-spheres.  This provides the homology relation needed
to satisfy the K\"ahlerity condition.  The resolved manifold is obtained
by explicitly blowing up the $\ICP^2$ within $\ICP^4$.

More generally, we can consider a conifold transition which begins
with a Calabi--Yau which acquires $N$ nodes along some multi-conifold
locus in the complex structure moduli space, such that there are $M$ homology
relations among the vanishing cycles, with each vanishing cycle involved
in at least one of the relations.
Due to the homology relations, any three-cycle which meets
one of these vanishing cycles must meet at least two of them.
We refer to such cycles as ``magnetic'' three-cycles (complementary
to the ``electric'' vanishing cycles).  Since there are $N-M$ independent
homology classes of vanishing cycles, there must be an $(N-M)$-dimensional
space of such
``magnetic'' cycles.  In our example with $N=16$ and $M=1$, if the vanishing
cycles are represented by $\gamma_1, \dots, \gamma_{16}$ such that
$\gamma_1+\cdots+\gamma_{16}$ is homologous to zero, then there
are $15$ ``magnetic'' cycles $\beta_1, \dots, \beta_{15}$ with
$\gamma_i\cdot \beta_j=\delta_{ij}$ for $i\le15$, and
$\gamma_{16}\cdot\beta_j=-1$.

The ``magnetic'' three-cycles remain three-cycles in the singular limit, but
once we blow the space up, each of these three-cycles becomes a three-chain
whose boundary is a combination of the shrinking holomorphic two-spheres
(the combination being determined by  which of the vanishing cycles the
three-cycle met).  This
process is illustrated in figure 1, in which a three-cycle (meeting
two homologous vanishing cycles) opens up into
a three-chain after the transition.  These three-chains provide  $N-M$
homology relations among the shrinking two-spheres, so that the areas of
those spheres generate
only $M$ independent new K\"ahler classes.  It is now clear why there must be
at
least two vanishing cycles initially: if there were only one, there
would be a dual ``magnetic'' cycle meeting that vanishing cycle, which
would become a three-chain after the transition whose boundary would
be the two-sphere, which would thus be homologically trivial.  But a
holomorphic two-sphere on a K\"ahler manifold can never be homologically
trivial.

\iffigs
\midinsert
$$\vbox{\centerline{\epsfxsize=3in\epsfbox{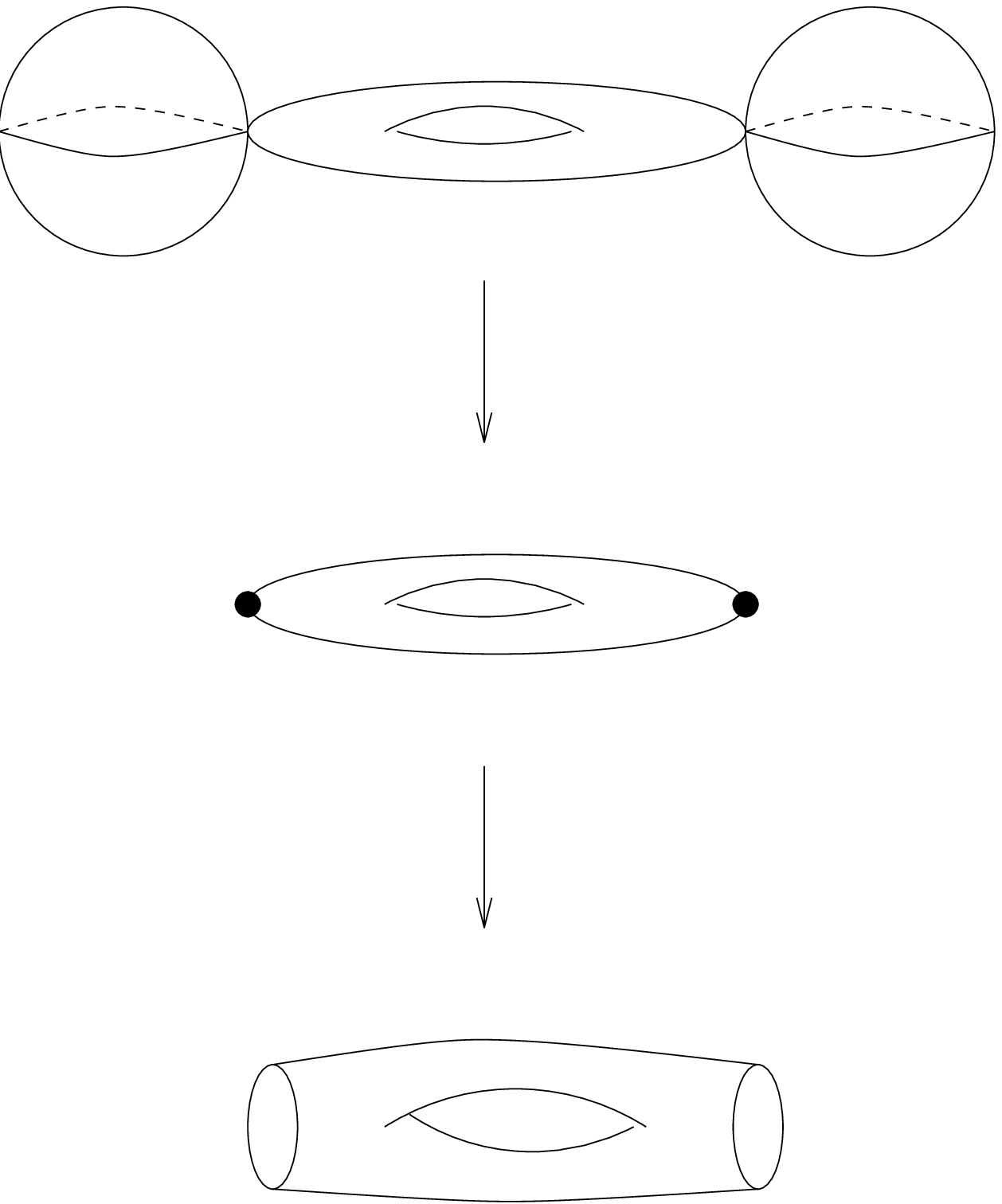}}
\centerline{Figure 1. A three-cycle becomes a three-chain.}}$$
\endinsert
\fi

Viewed from the opposite direction, we have $N$ shrinking
two-cycles with $N-M$ homology relations given by three-chains,
and $M$ dual ``magnetic'' four-cycles.  At the transition,
 the two-cycles have shrunken to nothing, and the the three-chains
have closed up to three-cycles; after the transition, there are new
``vanishing''
three-cycles dual to the three-cycles which came from three-chains,
and the old four-cycles open up into four-chains which specify the homology
relations among the vanishing three-cycles.

If we now consider the type IIB theory compactified on these Calabi--Yau
manifolds (as was done in \rGMS),
the vector-conifold transition begins with the variation of
complex structure to acquire $N$ nodes.  Due to the $M$ homology relations,
this multi-conifold locus occurs at complex codimension $N-M$ in the
complex structure moduli space.  Along that locus, the Dirichlet three-branes
which wrap the vanishing three-cycles become massless; that is, we get
$N$ new massless hypermultiplets at the transition.  The cohomology classes
which they span are associated with $N-M$ $U(1)$'s, and the $D$-terms
admit $M$ flat directions; going to the associated Higgs branch we find that
we have geometrically carried out the conifold transition.  The expectation
values for the surviving hypermultiplets correspond to the areas of
the holomorphic two-spheres---there are $M$ independent classes of these.

The Dirichlet three-branes wrapping the $N-M$ ``magnetic'' three-cycles remain
massive during this process (with some finite mass),
 and it is natural to ask what happens to them after the transition,
since the three-cycles have turned into three-chains.
In fact, this is the key question to ask if one wishes to understand the
hyper-conifold transition, and we will return to it in detail once we
have studied an analogous question in field theory.

\newsec{Some Field Theory Considerations}

In this section we consider some
aspects of the low energy field theory
description of conifold transitions whose
string theory counterparts will be the subject
of section 4.

In the previous section we have seen that in
the type IIB vector-conifold transition on the quintic
described in \rGMS\ we go to a locus in the quintic
vector moduli space where sixteen new hypermultiplets
arise from three-branes wrapped on degenerate
three-cycles. These sixteen hypermultiplets are
charged under fifteen $U(1)$ gauge factors, and it
is this unit numerical difference which is responsible
for the single flat direction,
in the low energy field theory description,
which  allows passage to
the new Higgs branch. Due to the flat direction, the Higgs branch
has an extra hypermultiplet, and it comprises
the hypermultiplet moduli space of $(86,2)$ Calabi--Yau, which
can be realized as a complete intersection in $\ICP^4 \times \ICP^1$.

For ease of discussion of this low energy field
theory process, we consider the simpler case of
two hypermultiplets $H^{(a)}$, $a=1,2$, each charged
under a single $U(1)$. We imagine that these
hypermultiplets arise from wrapping three-branes
around two degenerating three-cycles $A_1$ and
$A_2$ which satisfy the homology relation
$A_1 + A_2 = 0$. This implies that their charges, $q^a$
under the $U(1)$ are $1$ and $-1$ respectively.
Recall that each hypermultiplet contains two
complex scalar fields $h^{(a)\alpha}$, $\alpha=1,2$,
giving us a total of eight real scalar fields.
The relevant part of the $N = 2$ supersymmetric
Lagrangian describing these fields is
\eqn\eLOWE{|{\cal D} H^a|^2 +
D^{\alpha \beta}D_{\alpha \beta} + ... }
where the $D^{\alpha \beta}$ are
\eqn\eDs{ D^{\alpha \beta} = \sum_a q^a(h^{*(a)\alpha}
h^{(a)\beta} + h^{*(a)\beta} h^{(a)\alpha}),}
where $h^{*(a)\alpha} = \epsilon^{\alpha}_{\gamma} {\overline
h^{(a) \gamma}}$ with an overline denoting complex
conjugation,
and ${\cal D}$ is the gauge covariant
derivative with connection $A$.

Setting the scalar potential to zero involves
three real constraints $D^{\alpha \beta} = 0$,
together with one $U(1)$ gauge invariance. These
conditions collectively reduce the original configuration
space by what is known as a hyper-K\"ahler quotient.
Briefly, whereas a K\"ahler manifold is equipped with
a K\"ahler two-form, a hyper-K\"ahler manifold is endowed
with three such forms. Such manifolds always have
real dimension $4d$ for an integer $d$. If the latter space also admits
a $U(1)$ action under which these three two-forms are invariant,
there is a natural way to  construct a new hyper-K\"ahler space
with dimension $4(d - 1)$.
Following \rKron, each two-form gives rise to a moment map whose
vanishing locus  reduces the dimension of the space by one real unit,
and choosing a transverse slice to the gauge orbits yields the
fourth real constraint. It is not hard to show that the $4(d - 1)$
dimensional space which results inherits a hyper-K\"ahler structure
from that on  the original space.

In the present context, our original configuration space is the
hyper-K\"ahler space $\IC^4$ consisting of the four complex scalar
fields $h^{(a) \alpha}$. A convenient hyper-K\"ahler structure
arises from the three two-forms
\eqn\eformsa{dh^{(1)1}d\overline h^{(1)2} + dh^{(2)1}d\overline h^{(2)2}}
\eqn\eformsb{dh^{(1)1}d\overline h^{(1)1} + dh^{(1)2}d\overline h^{(1)2}
+ dh^{(2)1}d\overline h^{(2)2} + dh^{(2)2}d\overline h^{(2)2}  }
\eqn\eformsc{dh^{(1)2}d\overline h^{(1)1} + dh^{(2)2}d\overline h^{(2)1}}
and the natural $U(1)$ action arises from gauge transformations.
Applying the procedure indicated above,
the three moment maps are precisely the three $D$-terms in
\eDs\ and hence the hyper-K\"ahler quotient thus yields
the moduli space of gauge inequivalent vacuum configurations.

Concretely, this hyper-K\"ahler quotient space has
complex dimension two, which corresponds to the scalar
degrees of freedom in a single hypermultiplet. Thus,
the moduli space of vacua consists of
 a single flat
direction in the potential. Explicitly, this
flat direction arises from the four parameter
family of solutions (up to a $U(1)$ gauge
transformation)
\eqn\eflat{ |h^{(1)\alpha}| = |h^{(2) \alpha}|}
\eqn\eangles{ \theta^{(1) 2} - \theta^{(1) 1} =
\theta^{(2) 2} - \theta^{(2)1}}
where we have written
\eqn\enotation{h^{(a) \alpha} =
 r^{(a) \alpha} e^{i \theta^{(a)
 \alpha}}.}
A convenient solution to these conditions is the one chosen
in \rGMS, namely
\eqn\esolution{h^{(a) \alpha} = v^{\alpha}}
for any choice of the complex two-vector $v^{\alpha}$.
For nonzero values of $v^{\alpha}$ we therefore
spontaneously break the $U(1)$ gauge symmetry,
one hypermultiplet disappears via the Higgs
mechanism, and we are left with one net hypermultiplet.

We were careful to indicate above that the vacuum configuration
arising via such a hyper-K\"ahler quotient need not be a manifold---it
 can have singularities. There are two convenient ways to see this.
First, the choice we made in \esolution\ only partially fixes the
gauge freedom. Namely, $v^{\alpha}$ and $-v^{\alpha}$ are choices
in \esolution\ which differ by a $U(1)$ gauge transformation on our fields
with gauge parameter $\pi$. Thus, we need to impose a further
$\IZ_2$ discrete invariance on \esolution\ to have the set of
gauge inequivalent vacua. The vacuum configuration space is thus
$\IC^2/\IZ_2$. A second way of seeing this is to work directly
with gauge invariant combinations of our fundamental fields.
Namely, let $x = h^{(1) 1}\overline h^{(2) 2}, y = h^{(2) 1}\overline h^{(1)
2},
z = h^{(1) 1}\overline h^{(1) 2}, t = h^{(2) 1}\overline h^{(2) 2}$,
each of which is gauge invariant. These four combinations
manifestly satisfy the constraint $xy = zt$. Furthermore, the
constraint \eDs\ implies $z = t$. Thus, we arrive at the vacuum
space $\IC^3/(xy = z^2)$ which again is $\IC^2/\IZ_2$.

Whenever we Higgs a $U(1)$ gauge factor, we have
the possibility of generating the string-like
topological defects of Nielsen and Olesen%
---that
is, cosmic strings. At first sight, in the presence
of two hypermultiplets, one might think that we
would have the possibility of two kinds of cosmic
strings associated with nontrivial windings of each
of the two  fields. Indeed, this is true and each would
correspond to a global cosmic string, neither of which has finite
energy due to infrared divergences.  However, pairs of these
global strings where both fields undergo appropriate monodromy
at infinity can be reinterpreted as a finite energy cosmic
string of the Higgsed $U(1)$.  More explicitly, we take the hypermultiplets
to have the asymptotic
behavior
\eqn\ecosmic{h^{(a) \alpha} = C^{\alpha} e^{i q^a\phi}}
where we express our fields as functions of
cylindrical coordinates $(r,\phi, z)$ in $\IR^3$,
and the string points in the $z$-direction.
Finite energy per unit length requires
the gauge field $A_\mu$ to have the asymptotic
form
\eqn\egauge{A_{\phi} \sim {{1 \over r}}}
in order to cancel the infrared diverging
kinetic contributions to the energy.
This means
that
\eqn\eflux{\oint \vec A \, d \vec l = 1}
and hence the cosmic string carries a single unit
of magnetic flux.

The situation in such a configuration is that
the original $U(1)$ gauge symmetry is Higgsed
by the non-zero vacuum expectation values.
The only remnant of the original gauge symmetry
is the magnetic flux  confined in the core of
the cosmic string, where the massive vector multiplet
becomes massless. If for instance
we followed a magnetic monopole through the
symmetry breaking, we would find that its flux
becomes trapped into such tubes---a finite total
energy configuration can arise from following
a monopole/anti-monopole pair which can reside
at the endpoints of these cosmic flux tubes
after passing through the phase transition.

There are two relevant scales which set the size
of a cosmic string: the Higgs mass and the symmetry
breaking  induced vector meson mass.
One subtle feature in the field theory manifestation of
these ideas as they arise in the fundamental string context,
as we shall see in the next section, is that
the Higgs potential is exactly flat. It would
be interesting to study how field theory corrections
serve to capture all of the fundamental string
phenomena we will encounter in the next section, but
as we only use field theory as a qualitative guide,
we shall not do so here. Rather, the only extension
of the field theory discussion we shall need is
the generalization to multiple $U(1)$ factors, which
we now describe.

The generalization of this discussion to
any number of hypermultiplets charged under
a smaller number of $U(1)$ gauge groups is
immediate. For instance, in the case of
a transition like the vector-conifold transition
on the quintic in which there is one homology
relation between the $N$ vanishing three-cycles,
we will have
$N$ hypermultiplets $H^{(a)}, a = 1,...,N$
charged under a $U(1)^{N-1}$ gauge symmetry,
$U(1)_I, I = 1,...,N-1$'s with charges
\eqn\echarges{q^a_I = \delta^a_I}
for $1 \le a \le N-1$ and
\eqn\elastcharge{q^{N}_I = -1}
for all I.
The vacuum manifold is again found by minimizing
the potential---a procedure which as above
applies a hyper-K\"ahler quotient construction ($N-1$ times) to
the original hyper-K\"ahler $\IC^{2N}$ field configuration
space. Specifically, this
yields the conditions
\eqn\eflatgen{ |h^{(1)\alpha}| = |h^{(N) \alpha}|}
\eqn\eanglesgen{ \theta^{(1) 2} - \theta^{(1) 1} =
\theta^{(N) 2} - \theta^{(N)1}}
We have $4N$ real scalars, $3(N - 1)$ real constraints
and $N-1$ gauge invariances thereby leaving
a single hypermultiplet flat direction which can
be conveniently parameterized as in \esolution\ by
\eqn\esolutionnew{h^{(a)\alpha} = v^{\alpha}} where
$a$ now takes values from $1$ to $N$.
As in our simple two-field case discussed earlier, the vacuum
configuration has an orbifold quotient singularity at the origin.
As above, we can see this in two ways. First, the choices
$v^{\alpha}$ and $\omega v^{\alpha}$ where $\omega$ is a fundamental
$N^{th}$ root of unity are related by a discrete subgroup
of our original $U(1)^{N-1}$ symmetry; namely, the diagonal
element with gauge parameter $2 \pi/N$ for each $U(1)$ factor.
Second, we can again work directly with gauge invariant combinations
of our fields: $z_i = h^{(i) 1} \overline h^{(i) 2}$
$x = (\prod_{i = 1}^{N-1} h^{(i) 1}) \overline h^{(N) 2}$
$y = (\prod_{i = 1}^{N-1} \overline h^{(i) 2})  h^{(N) 1}$.
These manifestly satisfy the constraint
\eqn\econstgen{z_1z_2...z_N = xy}
and the $D$-terms equations imply $z_i = z_j$ for all $i$ and $j$, thereby
yielding the vacuum moduli space $\IC^3/(z^N = xy)$. From either
point of view, therefore, the moduli space is
$\IC^2/\IZ_N$.

As in our two-field case, in the Higgs
phase we can again form cosmic strings with the
same constraints as previously, giving us
$N- 1$ distinct string solutions, each carrying
a single unit of magnetic flux of one of the
$U(1)$ gauge groups. In this way, then, each of
the dual magnetic fields are confined into
vortex lines in the Higgs phase. Monopole/anti-monopole
configurations connected by these cosmic strings
provide the fate of free Coulomb phase magnetically
charged particles when the symmetry breaking phase
transition occurs.

In the next section we shall see this simple field
theory discussion realized through the geometry
of conifold transitions.

\newsec{Branes on Chains and Gauge Confinement}

In physical terms, the vector-conifold transition
is most conveniently described as a Higgsing
of a gauge symmetry.  If we wish to understand the reverse
transition---the hyper-conifold transition---we thus
have to understand un-Higgsing.  On the face of it this might
seem like an easy task;  all we have to do is
identify, in the Higgs phase,
a  vector multiplet which
has become massive by `eating' a hypermultiplet.
This however is not a simple matter because such a massive state
is not a reduced BPS supersymmetry multiplet and thus there is no
reason for its stability.  Thus we cannot identify it with any
stable particle or soliton in our string theory setup.
How else can we detect the existence of the Higgs phase?
In the Higgs phase we have a broken gauge symmetry and the
corresponding short range force due to electrically charged objects.
On the other hand, if we have massive magnetically charged objects in the
Higgs phase they will be confined and their magnetic flux
will be confined to a tube.
 These magnetic
flux tubes are simply the cosmic strings of the broken
local $U(1)$ gauge symmetries.
The existence of massless vectors and hypermultiplets---which are
generically massive in the Higgs phase---can
 be directly seen as propagating
massless modes moving up and down the cosmic string.

So the question to ask is whether we have any magnetically
charged states in our string theory context?  Indeed we do.
Let us for simplicity start with the case where we have one
$U(1)$ and two homologous $S^3$ vanishing cycles.  By a choice
of convention we call the three-branes wrapped around the vanishing $S^3$'s
electrically charged states with respect to this $U(1)$.
As discussed in section 2, there is a dual three-cycle which
intersects each of these $S^3$'s at one point.  If we consider
a three-brane wrapped around such a dual cycle we get a particle
which is magnetically charged with respect to the $U(1)$.
By choosing a minimal volume cycle in this dual homology class,
the wrapped configuration is a BPS magnetically charged state, $m$.
 Of course, in this un-Higgsed
phase what we call electric and magnetic is a matter of convention.
In particular, if we put a magnetically charged BPS state $m$
at $x$ in the uncompactified spacetime,
and its anti-particle state $\overline m$, which is the three-brane
wrapped with opposite orientation about the same three-cycle,
at $y$ then there is a Coulomb potential for this
configuration proportional to $1/|x-y|$. Let us
keep these magnetically charged states $m$ and $\overline m$
fixed at $x$ and $y$ respectively and pass to
the Higgs phase.  At this point the physics suggests
that the magnetic charges should get screened and hence
there should be a linearly growing energy for moving them
apart.  In particular the energy of the same configuration
should now go like $T |x-y|$
 where $T$ is the tension
of the cosmic string connecting them.
  Note that the total energy in this
configuration should also include the bare mass of the
magnetically charged particles.  If we call the bare mass $M_0$,
we expect that the energy for this configuration
should go like
$$E=2M_0+T|x-y| .$$
We would like to see how  this is
realized geometrically.

As discussed in section 2, in the Higgs phase the two $S^3$'s
are replaced by two $S^2$'s which are in the same homology
class.   This homology relation implies that there is a three-chain $C_3$ whose
boundary is the difference of these $S^2$'s:
$$\partial C_3= S^2_1 -S^2_2$$
where we have denoted the two $S^2$'s by $S^2_i$, with
$i=1,2$.
Moreover, in the Higgs phase, the magnetic three-cycle which intersects
the original $S^3$'s is
naturally identified with this three-chain.  On the other
hand we now face the dilemma that we {\it cannot}\/ wrap
a three-brane around a three-chain.  This of course is
similar to the statement that we do not expect to have
a free magnetically charged state in the phase where the magnetic
charge is confined. However we should be able to consider
$m(x)$---${\overline m}(y)$ configurations in this context.
Indeed there is a very natural geometric construction
which precisely does this.

\iffigs
\midinsert
$$\vbox{\centerline{\epsfxsize=4in\epsfbox{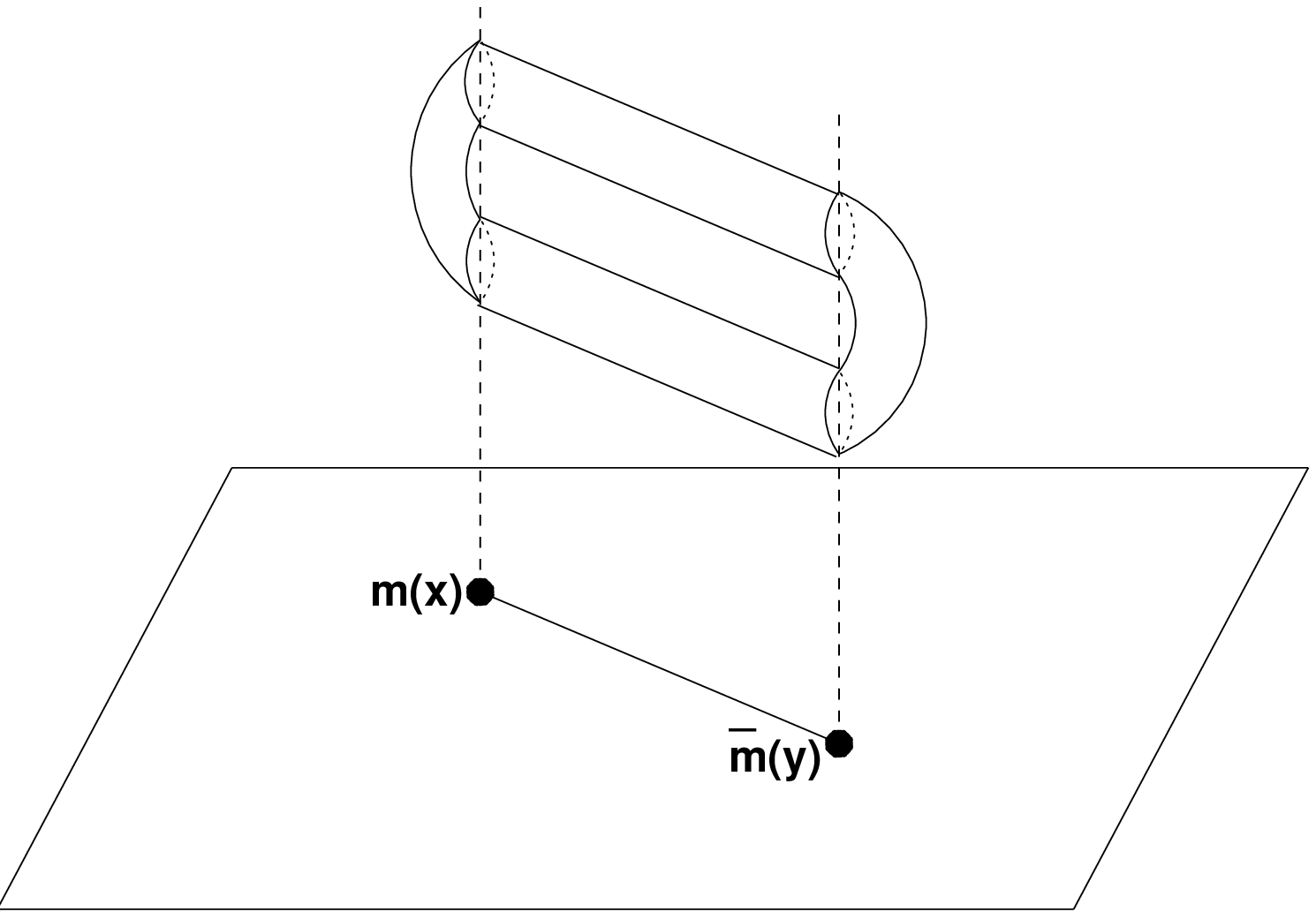}}
\centerline{Figure 2. Four three-chains assembled into a three-cycle,}
\centerline{yielding a confined $m\overline m$ pair connected by a flux
tube.}}$$
\endinsert
\fi

Consider the total space  $\IR^3\times X$
where $\IR^3$ is the uncompactified space and $X$ is the Calabi--Yau.
We will now construct a three-cycle in this total space
which will correspond to the configuration $m(x)$---${\overline m}(y)$.
  The three-cycle in question is
comprised of four three-chains which are joined together
along their boundaries (see figure 2, in which the base represents spacetime
and the vertical direction represents the internal space).  Let $I$ denote the
straight line from $x$ to $y$.
The four three-chains
are
$$m=x\times C_3, \quad f=
I\times S^2_1, \quad \overline m=y\times C^*_3, \quad {\tilde f}=I^*\times
S^2_2$$
where we have used $^*$ to denote the same chain with
the opposite orientation.
Note that the two chains we denoted by $m$ and $\overline m$,
which project to $x$ and $y$ in $\IR^3$ respectively,
are precisely the objects we would like to identify with
magnetically charged states $m$ and $\overline m$ at these
two points.  Moreover, the image of $f$ and $\tilde f$ in
$\IR^3$ is simply the straight line $I$ from $x$ to $y$.
Thus the image of this three-chain in $\IR^3$ is simply
two points $x$ and $y$ connected by a line $I$.  The identification
with the $m(x)$---${\overline m}(y)$ configuration suggests that the
$f\tilde f$ image on  $\IR^3$ should be viewed as the magnetic flux
of the cosmic string.  We will now
provide evidence for this identification of the magnetic flux tube.

In the Higgs phase, the hypermultiplet moduli
which take us away from the transition point can be viewed
as coming from the complexified K\"ahler deformations of the
Calabi--Yau as well as the moduli of RR gauge fields.
The RR gauge fields consist of an antisymmetric two-form
and an antisymmetric four-form field, which after decomposing with
a harmonic form on an internal $S^2$,  gives rise in the four
uncompactified dimensions
to a scalar and an antisymmetric two-form $C_{\mu\nu}$ respectively.
There is a cosmic string that $C_{\mu \nu}$ couples to.
In particular if we dualize $C$ to a scalar
$$dC=*d \phi ,$$
the cosmic string in question is characterized by
$\phi \rightarrow \phi+1$.  In fact it is easy to give
a geometric description of such a string. Since $C_{\mu \nu}$
came from the reduction of the four-form gauge potential, and that
couples to a Dirichlet
three-brane, the corresponding cosmic string is simply
a three-brane wrapped around the holomorphic $S^2$.  In the
example at hand we have two $S^2$'s and so we will get
two cosmic strings from wrapping the three-brane around
each of the $S^2$'s. These cosmic strings which are BPS saturated
do not have finite energy per unit length.
 This is in fact true
of any global cosmic string.\foot{
The situation here is the same as the seven-brane
of type IIB, which by itself will have infinite
energy,  However in the type IIB by utilizing the fact
that the dual scalar has $SL(2,{\bf Z})$  monodromy
we can get finite energy solutions by taking seven-brane
configurations which are relatively non-local
\vf\ by using the stringy cosmic
string geometry \scs.}
  In the Higgs phase the BPS saturated
strings correspond to global strings and not to cosmic strings of a broken
local gauge symmetry---the latter do not carry a BPS charge \ss.
There is an easy way to see that
the global cosmic string we are dealing with is not quite
the local cosmic string we are interested in, to which
we will now turn.

As discussed in section 3 it is useful to write
the hypermultiplet degrees of freedom in terms of composites of the
charged fields so that they are $U(1)$ neutral.
In particular, if $(h^{(1)1},{\overline h^{(1)2} })$ denote the hypermultiplet
degrees of freedom coming from one vanishing $S^3$ and
$(h^{(2)1},{\overline h^{(2)2} })$ those
coming from the other, we recall that
%
%
%
$x = h^{(1) 1}{\overline h^{(2) 2}}, y = h^{(2) 1}{\overline h^{(1) 2}},
z = h^{(1) 1}{\overline h^{(1) 2}}, t = h^{(2) 1}{\overline h^{(2) 2}}$,
and that relation $xy=zt$ becomes $xy=z^2$, after imposing the vanishing
of the potential.
As will be discussed in more detail in section 6, the
moduli space of hypermultiplets near this region is two `cosmic
strings' coming together at $z=0$, where the two cycles of the elliptic
fiber are identified with RR gauge fields on $S^2$ and a dual four-cycle.
In particular the RR gauge field corresponding to the vanishing
$S^2$ corresponds to going around $x,y$ in phase and keeping $z$
fixed:
$$x\rightarrow x \exp(i\phi), \quad
 y\rightarrow y\exp(-i\phi), \quad
z\rightarrow z .$$
In terms of the charged fields we can interpret this transformation
as either
%
%
$$h^{(1) \alpha} \rightarrow e^{i \phi} h^{(1) \alpha}; h^{(2) \alpha}
\rightarrow  h^{(2) \alpha}$$
or the same form with the two hypermultiplets exchanged: $1\leftrightarrow 2$.
Thus the BPS string we have obtained from wrapping the
three-brane around each of the two-cycles should correspond
to one of these two possibilities. This means that each of these
BPS strings can be viewed as a global Nielsen-Olesen string \rKleb.
But for a cosmic string of the Higgsed $U(1)$,
 {\it both}\/ hypermultiplets undergo phase rotations in such a way that
all gauge invariant objects are invariant and, in particular, $x,y,z$
undergo no monodromy.   Thus we identify the cosmic string
of the local $U(1)$ with the superposition of the BPS cosmic strings
coming from each of the $S^2$'s with opposite orientation.
This string, therefore,
carries no net BPS charge.  In fact the $f\tilde f$ string
in the above three-brane configuration is clearly the same as
this cosmic string, which we identify with the confined magnetic
flux.

It is also interesting to connect the classical mass formula
we derived for the confined magnetic states with the energy
of the three-brane configuration we constructed.  The total
energy of the three-brane wrapped around the three-cycle
we considered is clearly the volume of the three-cycle, which consists
of the volume of four three-chains.  The volume of the
two three-chains which correspond to the two magnetically
charged states are naturally identified with the
bare masses of the two particles. The volume
of the other two chains is $2|x-y|T'$ where $T'$ is
the area of each of the holomorphic $S^2$'s.  Noting
that two global strings give one cosmic string of the
local $U(1)$ and the
the tension of the global string is identified with $T'$,
we see that this is the same as $|x-y|T$.  We have thus
seen a simple explanation of the mass formula for the classical
magnetically charged states separated by a large distance.

Another interesting aspect is the issue of
 zero modes on the cosmic string. The counting
of such zero modes can be directly accomplished
in the D-brane picture. Explicitly, we note that on
a Dirichlet-three-brane world volume there is an
$N=4$ $U(1)$ gauge theory
\refs{\dlp,\witb} whose degrees of freedom
correspond to open string states with ends
lying on the three-brane.
Moreover, after wrapping the three-brane
around an $S^2$ in the CY we get a twisting of
this $N=4$ Yang--Mills \bsv, which has the same degrees
of freedom as the dimensional reduction of $N=1$ Yang--Mills
 in four dimensions down to two. Thus from each of the BPS
strings we get two massless bosonic and two massless
 fermionic modes and thus putting the two together we
get a total of four bosonic and four fermionic zero modes.
It would be interesting to understand this from a field theory
viewpoint.

Clearly, the discussion above can be generalized to many other
cases. For example if we have a $U(1)^{N-1}$ gauge group
which we identify with the Cartan subalgebra of $SU(N)$, and we have $N$
charged hypermultiplets corresponding to the fundamental weight
in the weight lattice of $SU(N)$, we can Higgs this system
and end up with one massless hypermultiplet.  The magnetic charges
will belong to the dual lattice, which is simply the root lattice
of $SU(N)$.  We expect to have $N-1$ basic types of cosmic strings
to be identified with $N-1$ broken $U(1)$ gauge symmetries.  In the
geometric description of the Higgs/confining phase we have $N$
homologous vanishing $S^2$'s.  There are $N$ basic BPS saturated
cosmic strings (they form the fundamental of $SU(N)$).  The
cosmic strings of the broken $U(1)$'s
are to be identified with the combination of BPS/anti-BPS
strings.  They form the adjoint weights of $SU(N)$, and they
are generated by $N-1$ elements, which we identify with $N-1$
basic cosmic strings of the higgsed $U(1)^{N-1}$.
The magnetically charged objects in the confining phase
correspond to the three-chains whose boundaries are the differences
of two homologous $S^2$'s and so they correspond to the
adjoint weights of $SU(N)$. Pairs of magnetically charged
states with opposite charge get connected with magnetic flux tubes
which are the corresponding cosmic strings carrying the adjoint
weights of $SU(N)$.

\newsec{Moduli Space Attachment}

In this section we give a more precise discussion of how the topologically
distinct Calabi--Yau moduli spaces connected by conifold transitions
actually attach. For concreteness, we shall focus on the quintic hypersurface
and the $(86,2)$ model, although our discussion is more general.
In fact, we still lack a first principles argument which ensures that
the vector-conifold transition discussed in \rGMS\ and reviewed
in section 2 does in fact take us to the Calabi--Yau space given by
the complete intersection of a bidegree $(4,1)$ and $(1,1)$ in $\ICP^4
\times \ICP^1$. There is essentially airtight circumstantial evidence
for this being the case: the confluence between the mathematics of
degenerate deformations followed by resolving small resolutions  and
the physics of degenerate points in Coulomb sector the moduli space
followed by a resolving Higgs mechanism, is very convincing. In this section
we add to this evidence through more global considerations which allow
us to explicitly see the attachment of the quintic K\"ahler moduli space
within the K\"ahler moduli space of the $(86,2)$ model. Furthermore,
and somewhat surprisingly, we see a dramatic shift in the phase boundaries
from the prediction of the linear sigma model to the reality of the
nonlinear model, which plays a key part in yielding a consistent
moduli space attachment.

To begin the discussion, let's recall that the vector-conifold transition
we are discussing most naturally arises for fixed and large K\"ahler
class on the quintic, where we trust conclusions drawn from low
energy field theory considerations. However, as BPS states are the central
participants in the transition, and since they are stable under local variation
of parameters, the transition can actually take place for a range of values of
the K\"ahler class on the quintic.\foot{We note that in changing the
K\"ahler class for
which we perform a vector-conifold transition, we may pass through
marginal stability curves for BPS states. Nonetheless, we expect
to be able to perform the transitions at generic points.} Thus,
we actually have a one-parameter family of vector-conifold transitions,
taking us into the two-dimensional K\"ahler moduli space of the $(86,2)$
model. The one-dimensional K\"ahler moduli space of the quintic therefore
must attach to a one-dimensional locus inside the K\"ahler moduli
space of the $(86,2)$ model.

To understand this attachment, let's first review the structure
of the K\"ahler moduli space of the $(86,2)$ model.
The realization of this model originally given in \texasiii\
is as a bidegree $(4,1)$ $(1,1)$ complete intersection in $\ICP^4 \times
\ICP^1$. For our purposes, though, there is an equivalent but more
convenient realization of this Calabi--Yau. Namely, the degenerate
complex structure we pass to in the quintic moduli space is of the form
\eqn\edegen{x_4 G_4(x_1,...,x_5) + x_5 H_4(x_1,...,x_5) = 0}
where
$(x_1,...,x_5)$ are homogeneous $\ICP^4$ coordinates with $G_4$ and
$H_4$ being homogeneous quadrics. The small resolution we perform
to pass to the $(86,2)$ model can be realized by blowing up
the $\ICP^2$ given by $x_4 = x_5 = 0$. Explicitly, we do this
by introducing additional homogeneous variables $y_1$ and $y_2$
satisfying $x_4y_1 = x_5y_2$. In this ambient toric space,
which is now a $\ICP^4$ blown up along a $\ICP^2$, the resolved
model is realized as a hypersurface. We can explicitly
realize this procedure torically by augmenting the toric data
for (the bundle ${\cal O}(-5)$ over) $\ICP^4$,
\eqn\etoricdata{\vbox{\centerline{$
v_0=(0,0,0,0,1),
v_1=(1,0,0,0,1),
v_2=(0,1,0,0,1),
v_3=(0,0,1,0,1),
$}\centerline{$
v_4=(0,0,0,1,1),
v_5=(-1,-1,-1,-1,1),
$}}}
by the additional point
\eqn\eaddpoint{v_6=(-1,-1,-1,0,1) = v_4 +v_5.}
In this toric variety, the equation of the Calabi--Yau,
obtained by forming the dual Newton polyhedron,
is precisely of the form
\eqn\edegenn{\sum_{(a,b)\ne(0,0)} x_4^a x_5^b x_6^{(a+b-1)}
F^{a,b}_{5-a-b}(x_1,...,x_3)= 0.}
Clearly, when the new variable $x_6$ is nonzero, we can use
one of the $\IC^*$ actions defining the toric
variety to scale it to one, in which case \edegenn\ becomes
\edegen. This establishes an isomorphism between the complement
of the singular locus on the quintic \edegen\ and the complement
of $x_6 = 0$ on \edegenn. When $x_6 = 0$, though, we see that
there are more points solving \edegenn\ than \edegen, reflecting
the fact that we have blown-up the singular locus.

The relation of this realization to the more standard
complete intersection form is immediate. Namely,
the ambient toric variety involves the
bidegree $(1,1)$ equation $x_4y_1 = x_5y_2$, which together
with the hypersurface equation \edegen\ yields the two
homogeneous equations in the complete intersection form.
Having established their equivalence, we shall henceforth
use the hypersurface realization for convenience.


The toric data \etoricdata \eaddpoint--- which is equivalent
to a linear sigma model formulation --- allows us, using the
methods of  \refs{\rAGM,\rWittenphases}, to describe the blown up
space in terms of seven fields and a $U(1)\times U(1)$ action
with charges given by
\catcode`\|=\active
\eqn\efieldcharges{\vbox{
\begintable
        |$x_1$|$x_2$|$x_3$|$x_4$|$x_5$|$x_6$|$p$\elt
$U(1)_1$|$1$  |$1$  |$1$  |$1$  |$1$  |$0$  |$-5$\elt
$U(1)_2$|$0$  |$0$  |$0$  |$1$  |$1$  |$-1$  |$-1$
\endtable
}}
\catcode`\|=12
There
are four phases in this model, which can be identified as the
original $(86,2)$ model,  a flopped Calabi--Yau counterpart,
a hybrid model, and the Landau--Ginsburg phase. The phase
diagram is shown in figure 3.

\iffigs
\midinsert
$$\vbox{\centerline{\epsfxsize=4in\epsfbox{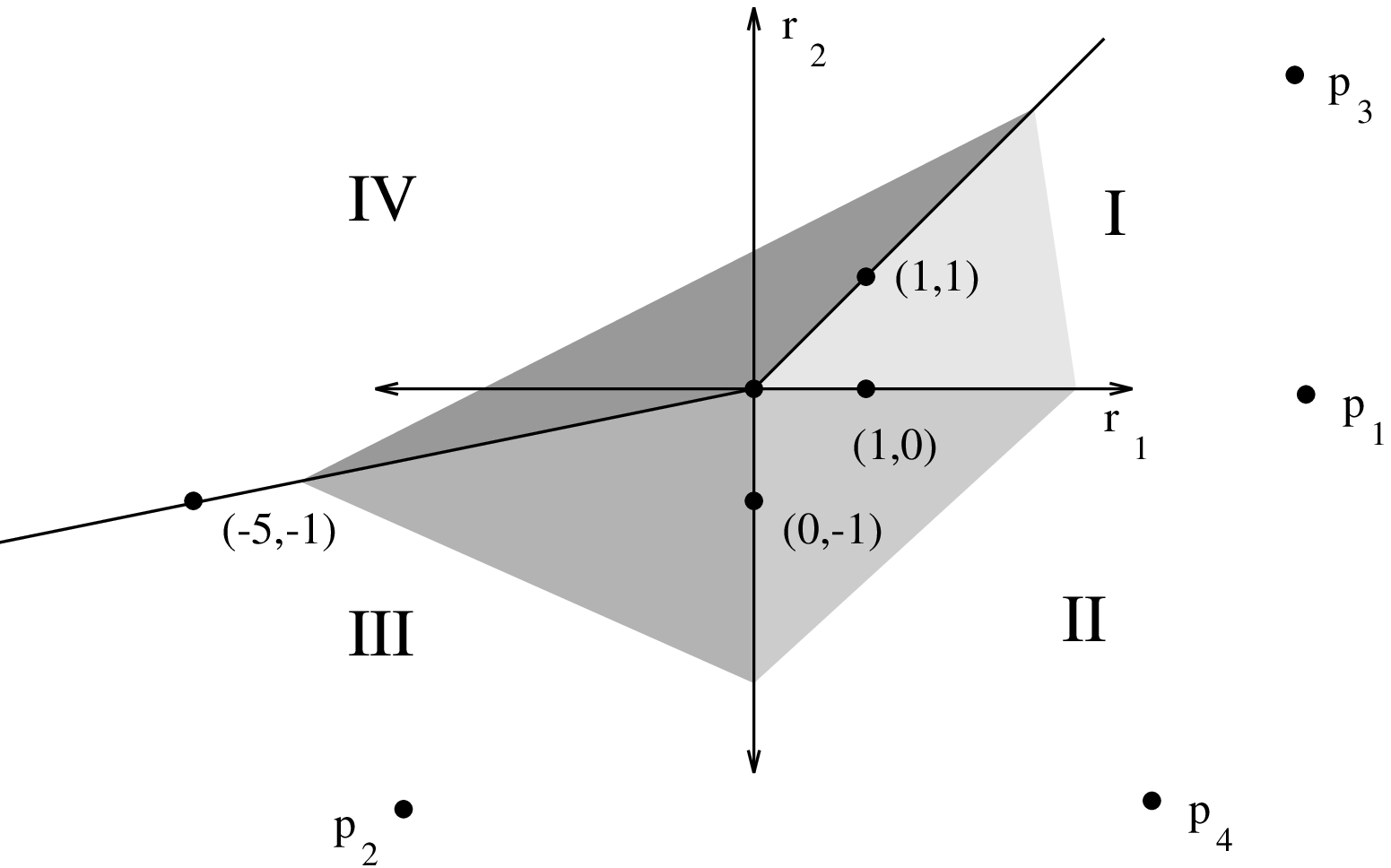}}
\centerline{Figure 3. The phase diagram.}}$$
\endinsert
\fi

To avoid confusion, we emphasize that this K\"ahler
moduli space is only the NS-NS base of the full hypermultiplet
moduli space; the full space includes RR toroidal fibers.

Where in this phase diagram do we expect the quintic K\"ahler
moduli space to attach? In the large radius limit
of the geometric part of the phase diagram, $r_1 \rightarrow \infty$,
we expect \rTrento\ that the quintic will attach at the center of the
flop---the place where the $S^2$ homology class, whose size varies by
the $r_2$ modulus, vanishes. In \rAGMsmall\ it was shown that in the
large radius limit, there are no stringy corrections to linear sigma
model coordinates associated with flops, and hence we would na\"{\i}vely
expect the attachment to pass through the point $p_1$, as indicated.
On the other hand, in the ``small radius limit'' of $r_1 \rightarrow -\infty$
(with $r_2$ held at $-\infty$),
we would expect the Landau--Ginsburg point of the quintic to attach to
the Landau--Ginsburg point of the $(86,2)$ model, at $p_2$. At first sight,
therefore, it is hard to see how the one-dimensional K\"ahler locus
of the quintic attaches to both of these points.

\iffigs
\midinsert
$$\vbox{\centerline{\epsfxsize=3in\epsfbox{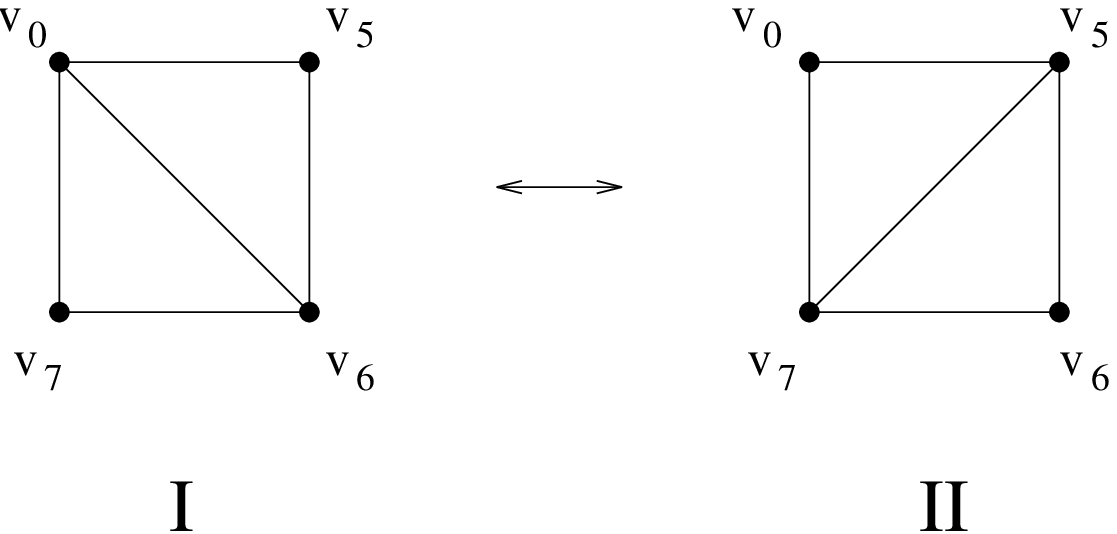}}
\centerline{Figure 4. The toric data for the flop.}}$$
\endinsert
\fi

The resolution to this problem requires a more careful study of the flop
transition taking us from phase I to phase II. Torically, this flop
transition is shown in figure 4. The key fact to notice is that this flop
makes us of the point $v_0=(0,0,0,0,1)$ which corresponds to the
anticanonical line bundle field in the linear sigma model,
represented  by the two-dimensional chiral field $p$ \rWittenphases.
Unlike the flops studied in \refs{\rAGM,\rAGMsmall}, therefore, the flop
we encounter here is not contained within the compact part of the
ambient toric variety, but makes use of the noncompact part as well.
We must therefore redo the analysis of \rAGMsmall\ to determine the
relationship between the true stringy volume of the $S^2$ and the linear
sigma model coordinate $r_2$.

To do so, we pass to the two-dimensional
complex structure moduli space of the mirror to  the $(86,2)$ model, which
can be obtained from the methods of \rBAT. Using the methods described in
\rAGMsmall, we can write down local coordinates valid in a neighborhood
of the two large complex structure points, mirror to points $p_3$ and $p_4$
in figure 3. As the moduli space is naturally expressed in the form
\eqn\etoricform{ (\IC^7 - F_{\Delta})/(\IC^{*5}) } with
$F_{\Delta}$ determined by the corresponding triangulation,
we can express the local coordinates in terms of $\IC^{*5}$ invariant
combinations of the initial $\IC^7$ toric coordinates, $a_0,...,a_6$.
 Written in this form, the coordinates are
\eqn\elocalcoords{z_{(I)1} = {{ a_1a_2a_3a_6 \over a_0^4}}, \quad
 z_{(I)2} = {{a_4 a_5 \over a_0 a_6}} },
\eqn\emorelocal{z_{(II)1} = {{a_1a_2a_3a_4a_5 \over a_0^5}}, \quad
z_{(II)2} = {{a_0a_6 \over a_4 a_5}} .}
{}From these we see that $z_{(I)2}$ and $z_{(II)2}$ are the local coordinates
for the rational curve in the moduli space connecting the mirror of points
$p_3$ and $p_4$. Having identified these coordinates, we can now write
down the Picard--Fuchs equation for the periods of this Calabi--Yau, along
the rational curve at large complex structure
(mirror to $r_1 \rightarrow \infty$). Following the methods
of \rBAT\  as applied in \rAGMsmall\ we find the following equation
\eqn\ePF{z(1-z) {{\del^2 f \over \del z^2}} + (1 - 2z){{\del f\over \del
z}} = 0}
for the periods  of the holomorphic three-form on the Calabi--Yau.
The two solutions to this equation with trivial and logarithimic
behavior near $z = 0$ will be denoted $f_0$ and $f_1$ respectively.
The important point to note is that this equation differs from eqn (63)
of \rAGMsmall\ due to the fact that this flop involves the noncompact
direction.
Thus, whereas the solution of the flop equation from \rAGMsmall\
gave
\eqn\enormalflop{ f_1 = \log(z), \quad f_0=\hbox{constant}}
and hence the mirror map
\eqn\emirrormapflop{ B + iJ = {{1 \over 2 \pi i}} \log(z)}
from which we see that there are no sigma model corrections, the situation
now is different.
It is immediate to solve \ePF\ and find
\eqn\enoncompactflopsoln{f_1 = \log{{z \over 1 - z}}, \quad
f_0=\hbox{constant}. }
This yields the mirror map
\eqn\emirrormapnoncompact{ B + iJ = {{1 \over 2 \pi i}}\log{{z \over 1 - z}}.}
We therefore see that there are strong sigma model corrections
near $z = 1$ in this case.

\midinsert
\catcode`\|=\active
\begintable
     |$z$     |$r_2$    |$z/(z-1)$|$J$    \elt
$p_3$|$0$     |$\infty$ |$0$      |$\infty$ \elt
$p_1$|$1$     |$0$      |$\infty$ |$-\infty$\elt
$p_4$|$\infty$|$-\infty$|$1$      |$0$
\endtable\catcode`\|=12
\centerline{Table 1.}
\endinsert

Let us now use this solution to measure the volume of the flopping $S^2$
curve in the $(86,2)$ model, with $r_1$ fixed at $ \infty$.
We directly see the results in table 1. Notice that
the location of the center of the flop $\int_{S^2}J = 0$
has undergone a dramatic shift from
the linear sigma model prediction and now resides on the lower toric
boundary component, at $p_4$. This means, recalling that the diagram of
figure 3 is just the real part of a complex phase diagram, that the
phase diagram at $r_1 \rightarrow \infty$ has undergone a shift
from the linear sigma model prediction.  That prediction,
 illustrated in figure 5, was that the boundary of the region of convergence of
the sigma model centered at $p_3$ would pass through $p_1$, so one might
expect that $p_1$ would be the center of the flop.  However, the shifted
prediction, illustrated in figure 6, is that the boundary of the region
of convergence actually passes through $p_4$ instead.
 We see that the phase boundary has substantially
shifted so that the two phases join along what appears to be the lower
horizontal boundary in figure 3. It is on this locus that the
string volume of the $S^2$ has shrunken to zero, not along the
locus $r_2=0$ of figure 3. This therefore resolves the previous puzzle,
as this locus also includes the respective Landau--Ginsburg points.

\iffigs
\midinsert
$$\vbox{\centerline{\epsfxsize=2in\epsfbox{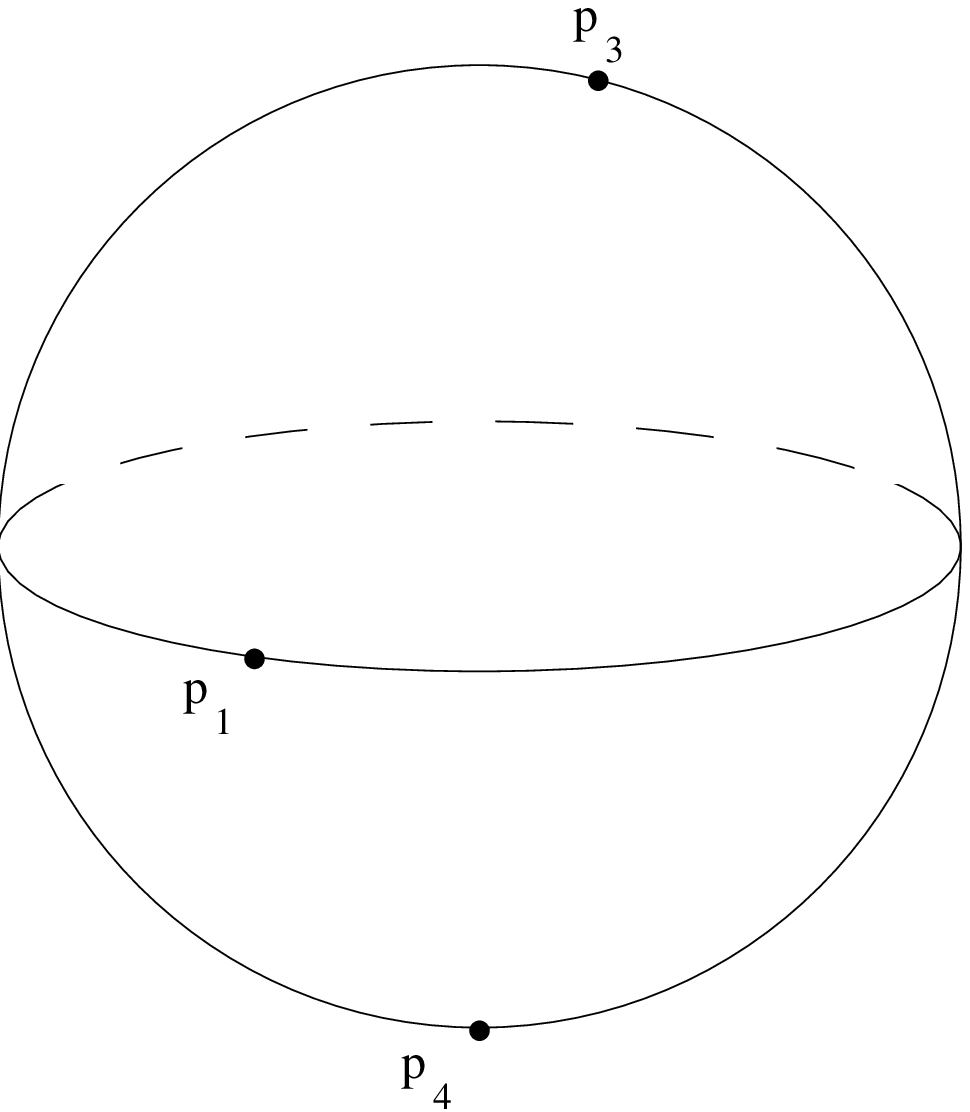}}
\centerline{Figure 5. Phase boundaries for the linear sigma model.}}$$
\endinsert
\fi

We can also quite directly see that this adjoining locus is manifestly
identified with the K\"ahler moduli space of the quintic. Again, it
is easiest to work in the mirror description where we intend to
identify this locus with the one-dimensional complex structure moduli
space of the mirror quintic.

\iffigs
\midinsert
$$\vbox{\centerline{\epsfxsize=2in\epsfbox{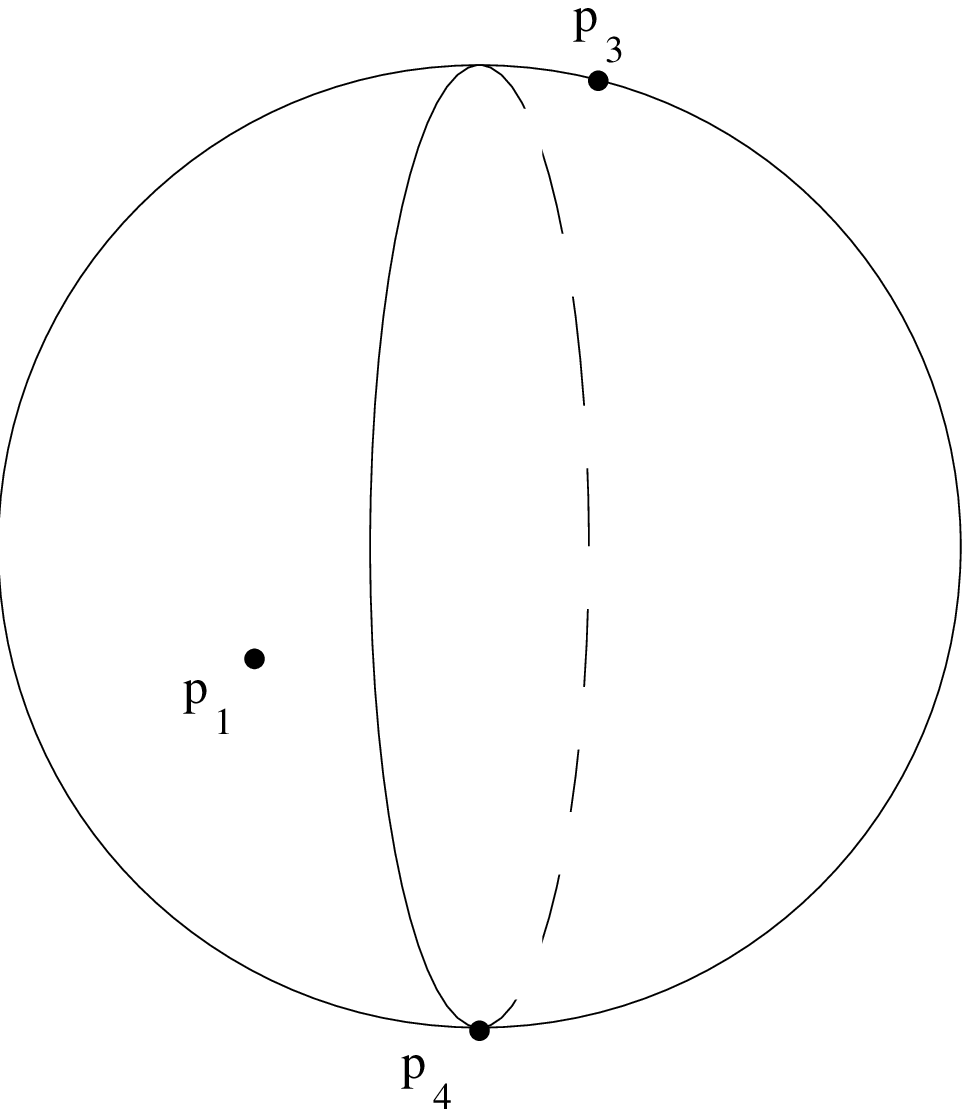}}
\centerline{Figure 6. Phase boundaries for the nonlinear sigma model.}}$$
\endinsert
\fi

To do so, let's recall that the complex structure moduli space of
the $(2,86)$ model, as implicitly indicated above, arises from
considering the possible coefficients in the defining equation.
If we wish to describe the equation completely, we would need $91$
homogeneous variables, which is a very cumbersome form to work with.
However, the essence of the defining equation is captured by the
equation for the hypersurface within $(\IC^*)^4$ (ignoring the boundary
of the toric variety).  In convenient coordinates $t_1, t_2, t_3, t_4$
that equation can be written
\eqn\eeqnoftwoeightysix{a_0+a_1t_1+a_2t_2+a_3t_3+a_4t_4+
a_5t_1^{-1}t_2^{-1}t_3^{-1}t_4^{-1}+a_6t_1^{-1}t_2^{-1}t_3^{-1}=0.}
If
we set $a_6 = 0$, we get precisely the corresponding equation for the
$(1,101)$ model:
\eqn\equintic{a_0+a_1t_1+a_2t_2+a_3t_3+a_4t_4+
a_5t_1^{-1}t_2^{-1}t_3^{-1}t_4^{-1}=0.}
To put this in a more familiar form, if we define
\eqn\ecoordchange{t_1=z_1^4z_2^{-1}z_3^{-1}z_4^{-1}z_5^{-1},\dots,
t_4=z_1^{-1}z_2^{-1}z_3^{-1}z_4^{4}z_5^{-1},}
then the equation takes the  form
\eqn\equott{z_1^{-1}\dots z_5^{-1}(a_0z_1z_2z_3z_4z_5+a_1z_1^5+\dots
a_5z_5^5)=0}
(of which  a $(\IZ_5)^3$ quotient must be taken since fifth powers occur
in the Jacobian determinant of the
change of variables; note that the factor of $z_1^{-1}\dots z_5^{-1}$,
which only affects the boundary of the toric variety, should be removed).

Thus, in this limit of $a_6 = 0$ we manifestly recover the one dimensional
complex structure moduli space of the mirror quintic.

To make contact with our previous discussion, let's recall that the
monomial-divisor mirror map of \refs{\rAGMmultiple,\rAGMmath} tells
us that the attachment locus $a_6 = 0$ corresponds
to the locus $r_2 = -\infty$, which
is precisely attachment locus found previously.

We have thus accurately found where in the moduli space the analysis of
the previous sections actually occurs.

\newsec{Stringy Quantum Corrections}

As discussed in detail in \refs{\rStrominger,\rBerkSie}, for type II
compactifications on
Calabi--Yau threefolds, even though the vector multiplets
do not receive quantum corrections, the hypermultiplet moduli space can.
We ask the question of whether there are such corrections
in our setup in the Higgs/confining phase.  Indeed as we will now argue,
the transitions we have
talked about {\it require}\/ that such corrections  exist. Moreover, we
will be able to say what these corrections are at least in the limit
where gravitational effects are ignored ($M_p\rightarrow \infty$).

The hypermultiplet states we have been studying
are mainly associated to the
K\"ahler moduli of a type IIB string on a Calabi--Yau manifold.
Equivalently, they correspond to the complex structure moduli
of  a Calabi--Yau manifold for a compactification of a type IIA string.
Let us focus on the latter interpretation.
If the Calabi--Yau has an $n$ dimensional complex structure moduli space,
then the hypermultiplet moduli space is a $4n+4$ real dimensional quaternionic
manifold.  This space consists of $2n$ real dimensions
parameterizing complex structure moduli, $2n+2$ real dimensions
counting the moduli of three-form gauge potential of type IIA
which we can associate with a $2n+2$ dimensional ``intermediate Jacobian''
torus corresponding to the periods of the holomorphic three form
of the Calabi--Yau, and two dimensions associated with the dilaton and axion
of type II strings. In more detail, the $2n$-dimensional base
${\cal M}$ specifying the complex structure moduli is the NS-NS
base discussed in the previous section. A useful bundle can be built
on this space whose fibers consist of $H^3(M_x,\IC)$ with $x$ denoting
a point in ${\cal M}$. A natural modification of this bundle, both
from mathematics and from physics, is to consider modding out
the fibers by integral shifts in cohomology $H^3(M_x, \IZ)$,
thereby yielding the so called bundle of intermediate
Jacobians, whose fibers are $H^3(M_x,\IC)/H^3(M_x, \IZ)$.
The latter, in fact, are complex tori as the Hodge decomposition
$H^3(M) = H^{3,0}(M) + H^{2,1}(M) + {\rm c.c.}$ yields
a natural complex structure on the fibers. Without the
axion-dilaton system, then, the hypermultiplet moduli space
is a bundle of complex $n+1$ (real
$2n + 2$) tori fibered over the $n$ complex
dimensional base. The axion-dilaton is included  as another complex scalar,
which
transforms in a complex line bundle over the moduli space,
thereby yielding the total $4n + 4$ dimensional quaternionic
moduli space.

 The perturbative metric on this
space has been discussed in detail in \refs{\CFG,\fer}.
We expect non-perturbative corrections to change the geometry
of this quaternionic manifold.
In principle there could
be two types of corrections: one type which is visible even if we
take the $M_p\rightarrow \infty$ limit and the other type consisting of these
gravitational corrections.  This is the case, for example,
with the $N=2$ Coulomb branch of the heterotic strings
where the field theory results \sew\
get embedded in string theory and can be isolated from the
further gravitational corrections \refs{\kv,\kklmv}.
In particular, we consider the limit in which  $M_p\rightarrow
\infty$ at the same time as we approach the conifold point.
Note that $M_p$ is related to the string coupling constant of type IIA theory
which we write as $S+\bar S$. In the limit
we are considering, we take $S+\bar S>>1$ at the same
time as we approach the conifold point.
In this connection it is helpful to recall that in the rigid
case the quaternionic manifold is replaced by a hyper-K\"ahler manifold.

A particular example where we expect there to be strong corrections
is for the type IIA theory near the conifold point where Euclidean membranes
wrapped around vanishing three-cycles are expected to deform
the geometry of the quaternionic
 moduli space \rBBS.  Let us review the singularity
of the hypermultiplet moduli in this context.  The most relevant
part of the singularity is associated with one complex modulus
of the Calabi--Yau, which we will denote by $z$,
 together with a two dimensional toroidal
subspace of the intermediate Jacobian, corresponding to the vanishing cycle
and its dual
cycle.  If we denote the complex structure of this torus by $\tau$,
the statement that we have a vanishing three-cycle at $z=0$ is, via
monodromy considerations,  the same as the
statement that the geometry of the singular space is described by
the elliptic fibration
\eqn\etorus{\tau(z) ={1\over 2\pi i}{\rm log}z .}
This description sounds very similar to the stringy cosmic string
considered in \scs\ where at $z=0$ there is a cosmic string.
Recall that in \scs\ a real two-dimensional torus is fibered
non-trivially over a one-complex dimensional base, and,
in the simplest case, the torus degenerates as in \etorus\ near
$z = 0$, and hence
we presently find ourselves in an analogous situation.
Actually more is true:
If we write the hyper-K\"ahler metric on this space using the
result \fer, the leading singularity
agrees identically with the metric found in \scs\ and the
K\"ahler form
given by
$$k=\tau_2 dzd{\bar z}+ \partial {\bar \partial}{(\zeta -\bar \zeta )^2
\over 2(S+\bar S)\tau_2} ,$$
where
$$\tau_2={\rm Im}\tau={1\over 4\pi}{\rm log}z\bar z$$
and $\zeta$ is a local holomorphic coordinate on the torus with
$\zeta\sim\zeta+1\sim\zeta+\tau$.
As discussed in \scs, this metric is hyper-K\"ahler everywhere, however its
Riemann tensor has singularities at $z=0$, which is the point
where the string theory perturbation theory breaks down.  Note that
the K\"ahler class of the fiber torus is $1/(S+\bar S)$ and so
in the weak coupling limit it goes to zero.

The case mainly dealt with in this paper, namely type IIB
with $N$ homologous simultaneously vanishing two-cycles,
 corresponds to type IIA with $N$ homologous simultaneously vanishing
three-cycles. In this case the relevant singular part of the
hypermultiplet moduli is again the same as given above, except
that now
\eqn\etau{\tau={N\over 2\pi i}{\rm log}z.}
This is a configuration analogous to $N$ stringy cosmic strings which have
come together
at $z=0$.

The mathematical version of this degeneration follows immediately
from the intermediate Jacobian formalism introduced above.
Namely, let's call the $N$ homologous vanishing three-cycles
$\gamma_1,..,\gamma_N$. Consider encircling the locus in ${\cal M}$
where this degeneration occurs. By the theorem of \lefschetz,
if $\gamma'$ is any other three-cycle, it experiences the monodromy
transformation
\eqn\emonodromy{\gamma' \rightarrow \gamma' + \sum_{i=1}^N (\gamma' \int
\gamma_i) \gamma_i }
upon traversing such an encircling path. Now, since all
of the $\gamma_i$ are homologous, if $\gamma'$ is a dual cycle to
the common homology class $[\gamma]$ of the $\gamma_i$, \emonodromy\ implies
that
\eqn\edual{[\gamma'] \rightarrow [\gamma'] + N[\gamma].}
Again, therefore, we see that this monodromy relation implies
that the local form of $\tau$ must be as in \etau.

{}From the field theory analysis, which should give
an accurate picture of the hypermultiplet moduli
in the limit in which we turn off gravity, we expect the relevant
singularity of the space be $\IC^2/\IZ_N$.  The question we
now wish to address is how this fits with the perturbative
description
of this space given above. This is relatively clear in that
as $N$ cosmic strings come together, the corresponding family\foot{
We use the ``Weierstrass model'' for this family, in which
the size of the torus is the only allowed K\"ahler parameter.}
of tori acquires  an
$A_{N-1}$ singularity in the fiber at $z=0$. As discussed in \scs, if we
perturb the metric to give the K\"ahler class of the fiber a finite
size, there is a unique Ricci-flat hyper-K\"ahler representative
of the perturbed metric; this will reproduce the expected singularity
behavior $\IC^2/\IZ_N$.  This differs from the metric we gave at
tree-level (with $(S+\bar S)\to \infty$), but the corrections needed
to reproduce a metric of the form $\IC^2/\IZ_N$
are uniquely determined by the value of the K\"ahler
 class of the
fiber which we identify with $1/(S+\bar S)$.  We conjecture
that this is exactly the quantum string corrected metric.
This in particular implies that the singularity of the
metric which appeared for $z=0$ along the whole torus
for $N>1$ is replaced by a point in the fiber with $\IC^2/\IZ_N$
singularity in accord with field theory expectations.
Note that at very weak coupling the exact metric becomes
arbitrarily close to the stringy cosmic string metric.
In this limit the hyper-K\"ahler metric of K3 {\it becomes}\/
the stringy cosmic string metric.  This actually is very much
the sense in which F-theory on K3 is connected with M-theory
on K3 upon
compactification on a circle \vf.

Even though we did not directly study the $N=1$ case
in this paper, i.e. the type IIA near a conifold singularity,
given the fact that the Euclidean membrane
corrections are of the same type \rBBS , we are led to
conjecture that  we get the same resolution in this case as well.
Note however that in this case the actual moduli space
will have no left over singularity.\foot{
The quantum corrections we have found in this
four-dimensional string theory are
similar to the ones encountered in certain three-dimensional field theory
systems recently studied \swt.}

In fact we can motivate this discussion in another way.
We use the results in \refs{\ov,\bsvii}
to replace the conifold in type IIA given by
$$xy=z, \quad uv=z-\mu$$
with type IIB with a five-brane at $z=\mu$ which is wrapped around
the elliptic fiber given by $x,y$ over the $z$ plane. Now we use the
fact that at strong coupling the symmetric five-brane is equivalent
to Dirichlet five-brane of the D-string.  If we dualize twice
on the fiber of the $T^2$ the moduli problem at hand is
given by a three-brane in the geometry $xy=z$. The moduli of this
theory should be identified with positions of the three-brane,
which can move along the space $xy=z$;  the smooth metric is the same as
that on the
$xy=z$
 fibration,
which has a manifestly non-singular hyper-K\"ahler metric as we
discussed above.

As mentioned above this is the leading correction
in the limit of ignoring gravitational effects. When
we turn on gravity we do expect that the qualitative features
we have found will persist but the actual metric will get corrected
in a non-universal way, depending on how the vanishing cycle
sits in the Calabi--Yau,
and the total space will be a quaternionic manifold.

\newsec{Extensions and Speculation}

It may appear that
the geometric description of confinement we have found
only corresponds to abelian confinement. However, we should
recall that in the work of \sew\ the confinement
of $N=1$ $SU(2)$ gauge theory can be continuously connected, upon
adding a massive adjoint, to the Higgsing of the magnetic
charge.\foot{The picture of confinement we present here
was initiated during conversations with Nathan Seiberg.}
In fact we believe the geometric picture we have found fits
well with the geometrization of the Seiberg--Witten system
found in \klmvw.  In particular, it was
found that  the BPS states of the $N=2$ Yang--Mills theory
correspond to wrapping of one-branes around cycles of a Riemann
surface.  This one-brane itself can be viewed as part of the
three-brane of type IIB partially wrapped around two-cycles in the
internal space.  The electric BPS states correspond to wrappings
of this one-brane around the A-cycle and magnetic BPS states to wrappings
around
the B-cycle.  In the limit where the B-cycle vanishes
one gets massless magnetic monopoles. If we add the mass deformation
to the adjoint breaking
$N=2\rightarrow N=1$ it is natural to expect that the Seiberg--Witten
torus opens up around the pinched point.
Following the discussions of this paper,  the electrically
charged BPS states are confined as they now correspond to wrappings of
one-chains. In particular,
pairs of them should have the same confining description
as we have found above---namely, one-branes which wrap
homologically trivial one-cycles in the total space, with the part of
the one-brane passing through uncompactified space
 being identified as the flux tube between the electrically
charged states  (see figure 7,
 in which once again the base represents spacetime
and the vertical direction represents the internal space)).

\iffigs
\midinsert
$$\vbox{\centerline{\epsfxsize=4in\epsfbox{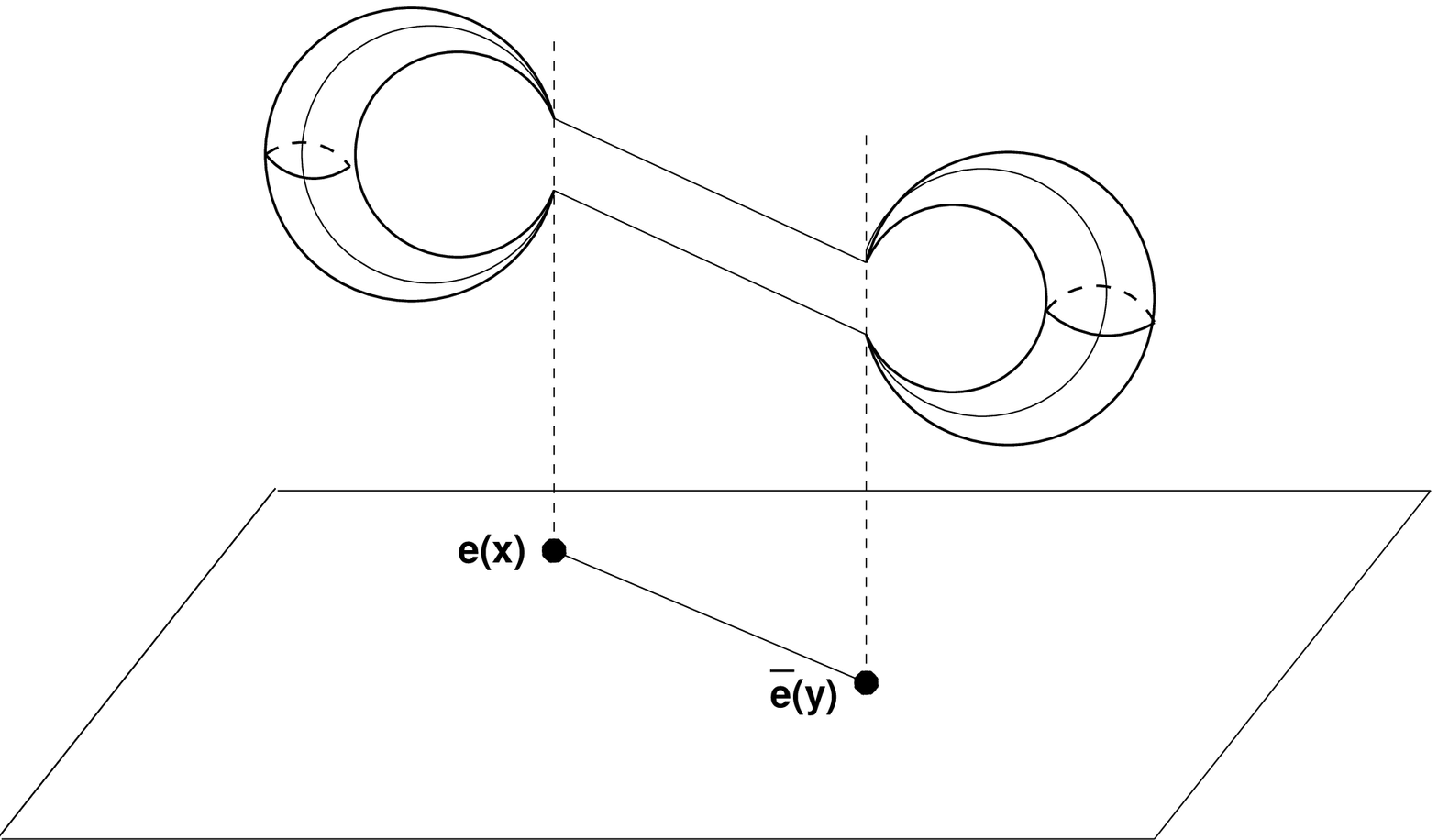}}
\centerline{Figure 7. A Seiberg--Witten torus opening up to a sphere,}
\centerline{with one of its one-cycles becoming a chain.}}$$
\endinsert
\fi

We can also extend the discussion of magnetically charged
states we have found beyond the cases which lead to
topological transitions.
Namely, if we wrap D-branes around chains which end on supersymmetric
cycles (which in principle can have more than two boundaries)
we can identify them with charged states in the confining
phase. In principle there are a large number of them in Calabi--Yau
compactifications.  For example if we consider
type IIB on quintic threefold there are generically
2875 degree one holomorphic curves. Since they are all in the
same homology class we can connect pairs of them with three-chains
 and identify these with charged states in the
confining phase.  Or, for example, we can have a three-chain
which ends on three two-cycles, two corresponding to degree one maps
and one corresponding to a degree two rational curve, with
appropriate orientations.

Note that even though we have mainly concentrated on electric/magnetic
particles
in this paper, the same considerations apply to higher $p$-branes.
In particular when the $p$-brane is in the Higgs phase the dual
$(d-p-4)$-brane is confined
 \rQT\ and this can be geometrically realized along the lines
we have discussed.  In particular,
  the confined $p$-brane will correspond to a $(p+r)$-brane wrapped
on an $r$-chain. An interesting example of this arises in compactifications
of M-theory on Calabi--Yau threefolds down to five dimensions. In fact
this would be related to what we have studied in the strong coupling limit
of type IIA theory on the same Calabi--Yau. If we consider type IIA
where some homologous $S^3$'s vanish, by wrapping the M-theory
five-brane around four-chains whose boundaries are vanishing $S^3$'s
we get strings in five dimensions which are confined by nearly tensionless
membranes flux sheets corresponding to wrapping pairs of five-brane around
pairs of vanishing
$S^3$'s.

It is also possible to study other types of Coulomb/Higgs
transitions along these lines.  In particular the
transitions involving the small $E_8$ instantons in connection
with vanishing $E_8$ del Pezzo in CY should be extremely interesting
to study in connection with the considerations of this paper.

\bigskip

\centerline{\bf Acknowledgments}

We would like to thank K. Becker, M. Becker, P. Candelas, M. Gross, K.
Intriligator,
Y. Kantor, N. Seiberg,
A. Strominger and A. Vilenkin for
valuable discussions.
We gratefully acknowledge the hospitality of the Aspen Center for
Physics, where much of this work was completed.
The research of B.R.G. is supported in part by the National Science Foundation,
the Alfred P. Sloan Foundation, and a National Young Investigator Award,
that of D.R.M. is supported in part by NSF grant DMS-9401447,
and that of
C.V. is supported in part by NSF grant PHY-92-18167.

\listrefs

\end